	\documentclass[10pt,journal]{IEEEtran}
\usepackage[usenames,dvipsnames]{xcolor}

\usepackage{cite}
\usepackage{amsmath,amssymb,amsfonts,graphicx,bbold,amsthm,mathtools}
\usepackage{enumitem}
\usepackage{subfigure,epstopdf,algorithm}
\usepackage{array}
\usepackage{xcolor}
\usepackage{thmtools}

\IEEEoverridecommandlockouts
\usepackage{times}
\usepackage{eqparbox}
\usepackage[noend]{algpseudocode}

\makeatletter
\def\BState{\State\hskip-\ALG@thistlm}
\makeatother

\begin{document}
	
	%\title{Kernel Regression for Signals over Graphs \\ Graph Signal Prediction using Kernel Regression where Input is Agnostic to Graph}
	
	\title{\textcolor{black}{Predicting Graph Signal{\color{black}s} using Kernel~Regression where {\color{black}the} Input Signal is Agnostic to {\color{black} a} Graph}}
	
	\author{Arun~Venkitaraman, Saikat Chatterjee, Peter H{\"a}ndel\\ 
		Department of Information Science and Engineering \\                   
		School of Electrical Engineering and Computer Science\\            
		KTH Royal Institute of Technology,  
		SE-100 44 Stockholm, Sweden                 \\
		arunv@kth.se, sach@kth.se, ph@kth.se
		%\thanks{The authors would like to acknowledge the support received from the Swedish Research Council.}
	}

	\markboth{}
	{}
	\maketitle
	
	\begin{abstract}
	\textcolor{black}{
	We propose a kernel regression method to predict a target signal lying over a graph when an input observation is given.
	 The input and the output could be two different physical quantities. In particular, the input may not be a graph signal at all or it could be agnostic to an underlying graph. 
	%In our method, the input signal used for prediction and the target signal to be predicted could be completely two different physical signals, for example, pressure and temperature. We treat the input signal as an observation that may not be a graph signal at all, or that it is agnostic to any underlying graph.
	  We use a training dataset to learn the proposed regression model by formulating it as a convex optimization problem, where we use a graph-Laplacian based regularization to enforce that the predicted target is a graph signal.  Once the model is learnt, it can be directly used on a large number of test data points one-by-one independently to predict the corresponding targets. {\color{black}Our approach employs kernels between the various input observations, and as a result the kernels are not restricted to be functions of the graph adjacency/Laplacian matrix.}
%	  and using the {\color{black} assumption} that the target vector is a smooth signal over an underlying graph. The  {\color{black} condition} is imposed using a graph-Laplacian based regularization. 
		%We show that the predicted target using the optimal regression coefficients for any given input is a function of a kernel of the input and the graph-Laplacian. 
		We show that the proposed kernel regression exhibits a smoothing effect, while simultaneously achieving noise-reduction and graph-smoothness. We then extend our method to the case when the underlying graph may not be known apriori, by simultaneously learning an underlying graph and the regression coefficients. 
		%We validate our theory by application to various synthesized and real-world graph signals. 
		Using extensive experiments, we show that our method provides a good prediction performance in adverse conditions, particularly when the training data is limited in size and is noisy. 
		%\textcolor{black}{Further, we experiment with real-world cases where observed input and predicted output are same physical quantities, and lying jointly over a graph. For such input-output data, we compare our method with state-of-the-art kernel ridge regression method of \cite{}.} 
In graph signal reconstruction experiments, our method is shown to provide a good performance even for a highly under-determined subsampling.
	}	
		
		%We consider the problem of regression for signals over graphs. By treating the graph signal as the target vector associated to an input vector, we learn the optimal linear regression coefficients from multiple training observations with the awareness of the underlying graph. We achieve this by imposing graph-Laplacian based smoothness constraint during the training process. We show that the target predicted using the optimal regression coefficients for any given input depends only on the kernel of the observations and the graph-Laplacian. This motivates us to propose kernel regression for signals on graphs through a dual representation or {\em kernel-trick}. We discuss how the regression output exhibits a smoothing effect simultaneously achieving noise-reduction and graph-smoothness. We further extend the kernel regression approach to simultaneously learn the underlying graph from the training signals. We validate our theory by application to various synthesized and real-world graph signals. Our experiments show that kernel regression over graphs outperforms conventional regression consistently, particularly for small sample sizes and noisy training. We also observe that regression reveals the structure of the underlying graph even using small number of training samples.
	\end{abstract}
	
	\begin{IEEEkeywords}
		Linear model, regression, kernels, machine learning, graph signal processing, graph-Laplacian.
	\end{IEEEkeywords}
	\begin{center}
		EDICS$-$ADEL-SIPOG
		%NEG-SPGR, NEG-ADLE, MLR-GRKN.
	\end{center}
	\IEEEpeerreviewmaketitle
	\section{Introduction}
	
	Graph signal processing (GSP) has emerged recently as a framework which employs graph-structural information in the analysis and processing of vector-valued signals \cite{Shuman,Sandry1}. The framework has been shown to exhibit great potential in a wide range of real-world applications that deal with data over networks or graphs. By actively making use of the graph or the network structure, GSP deals with the extension of several traditional signal processing and machine learning concepts to a graph signal setting. In this article, our contribution to GSP is the development of a {\color{black}supervised} kernel regression method for predicting graph signal outputs {\color{black}from general input observations}. In the next two subsections, we provide a review of the existing literature, followed by our contributions placed in the context of the existing methods.

\subsection{Literature review}	
The extension of traditional signal processing methods to graph signal processing {\color{black}includes} many conventional spectral analysis concepts such as the windowed Fourier transforms, filterbanks, multiresolution analysis, and wavelets \cite{Sandry1,Sandry2,Sandry3,Shuman, windowedGFT,Narang2010,Narang2012,Narang2013,Coifman2006,Ganesan,Hammond2011,Wagner2, vertexfreq,pyramidgraph, pp_graph1,pp_graph2,ArunSampta15,ArunGHT}.
	 Spectral clustering approaches based on graph signal filtering have also been proposed \cite{Tremblay,Tremblay2}. 
	%The algorithm is shown to have similar performance as the classical spectral clustering while being fast and scalable. 
	The problems of sub-sampling and interpolation of signals lying over graphs have been considered extensively in diverse settings\cite{chen2,graphsamp1,graphsamp2, graphsamp3,graphsamp4,graphsamp5,graphsamp6,anis,graphsamp10,graphsamp11,KRG_R1}. Techniques for compression and representation of signals such as the principal component analysis (PCA) \cite{graphPCA1,graphPCA2} and dictionary learning approaches \cite{Thanou2014, graphdict1,graphdict3, dualgraphelad} have also been proposed for graph signals. Many researchers have considered the statistical analysis of graph signals particularly in the context of stationarity \cite{statgraph1,statgraph2,statgraph2_journal,statgraph3,statgraph4,statgraph5}. The  reconstruction and estimation of graph signals have also been steadily gaining interest in the community. {\color{black} Berger et al. \cite{Berger17} and Chen et al. \cite{chen1} considered the recovery of graph signals based on a total-variation minimization formulated as a convex optimization problem. Wang et al. \cite{KRG_R2} considered a distributed reconstruction of time-varying bandlimited graph signals. {\color{black}Di Lorenzo} et al. \cite{KRG_R3} proposed a least mean squares approach for the adaptive estimation and tracking of bandlimited graph signals. Several approaches have also been proposed for learning an underlying graph structure from the given graph signal data \cite{Dong:2016, graphlearn8,graphlearn2,graphlearn3,graphlearn4,graphlearn5,graphlearn9,Chepuri_laplacian,SR7,graphrecon,gl_krg}. GSP is a rich and continually expanding area of research and we refer the reader to \cite{gsp_overview_ortega} for a more comprehensive review of the developments.

We now proceed to briefly survey the relevant literature in kernel regression and kernel methods. Kernel regression constitutes one of the fundamental building blocks of supervised and semi-supervised learning strategies, be it in simple regression tasks or in the more advanced settings. Kernel regression lies at the core of support vector machines \cite{SVM} and Gaussian processes \cite{GP_Seeger}, and finds applications in deep neural networks\cite{deeplearning, kernel_deeplearning} and extreme learning machines\cite{ELM_kernel1,ELM_kernel2}. 
	Kernel regression in the setting of graphs or manifolds has been investigated in the  labeling and coloring of graphs and in the context of graph clustering \cite{diffusionkernels,Smola2003,graphlabel2,graphlabel3,graphlabel4,graphlabel5,graphlabel6,graphlabel7}. These works generally deal with signals which are binary-valued. {\color{black}Kernel regression has been employed in image deblurring by using a graph-based constraint on the pixel intensities of the deblurred image\cite{takeda2008deblurring}. Kernel regression was recently used in object saliency detection and spatial attention modeling in images, wherein the kernel matrix was simultaneously used to define a Laplacian matrix, in order to recover the smooth images\cite{dou2017object}. These prior graph-based approaches incorporate a graph-Laplacian based regularization by defining a graph between the various observations/datapoints and are concerned with an output/target that is scalar valued, such as the node label or the pixel intensity.} {\color{black} Kernels have also been extensively employed in the smoothing and regression of brain signals, where the functional connectivity or the topology of the brain surface is described using meshes\cite{kerbrain_5,kerbrain_1,kerbrain_6,kerbrain_2,kerbrain_3,kerbrain_0}.} 
	%In contrast, our approach deals with predicting a graph signal for each individual observation by using the graph-Laplacian of the graph over which each target/output lies over and not a graph between the input observations as pursued by the prior works.} 

	Kernel-based reconstruction strategies specific to graph signals were proposed recently by Romero et al. in the framework of reproducing kernel Hilbert spaces\cite{kergraph1,kergraph2}. Using the notion of joint space-time graphs, Romero et al. have also proposed a kernel based reconstruction of graph signals and an extension of the Kalman filter for kernel-based learning \cite{kergraph3,kergraph4}. Along the same lines of thought, Ioannidis et al. proposed a more general approach for inferring functions over graphs in both static and dynamic settings\cite{IOANNIDIS2018173}. {\color{black}Kernel regression combined with diffusion wavelets have been employed in the modeling of mandible growth in CT images\cite{CHUNG201563}.} Shen et al. used kernels in structural equation models for the identification of network topologies from graph signals\cite{baingana1}.  The prior works of \cite{kergraph1,kergraph2,kergraph3,kergraph4,IOANNIDIS2018173} use kernels across the nodes of a graph, while considering the input comprising the signal values over a subset of the nodes of the graph. {\color{black}In these prior works, the setting is that all the observed inputs and the corresponding outputs to be predicted lie jointly over a composite or an augmented graph. As a result, the setting results in large-sized graphs which may not provide a scalable solution when the number of inputs and outputs becomes moderately large. Further, the setting naturally requires that the input and the output variables are of the same physical quantities. 
		%For example, in a scenario such as predicting the temperature of some of the cities in a country from the temperature observed over the remaining cities.
	Therefore, these prior works suffer from limitations when the input is a fundamentally different physical quantity from the predicted output, or when the input is not a graph signal. For example, consider a scenario where we observe the air pressure of several cities in a country as the input, and the task is to predict the temperature of those cities as the output.}
	
\subsection{Our contributions vis-a-vis existing works}

{\color{black}We propose a kernel regression method for graph signals that can handle scenarios where the input and the output may be entirely different physical quantities, or when the input is not a graph signal or is agnostic to a graph. For example, our method is applicable to the case when the hourly air pressure measurements over the cities in a country is taken as the input, and the predicted output is the temperature of those cities. (Such a real signal example is indeed demonstrated in the numerical experiments section.) This is possible because we treat the input variables without any graph constraints, or as being agnostic to a graph. The graph-awareness is employed only for the output: that the predicted output is a vector lying over a graph. Since we do not use kernels between the nodes of the underlying graph but only across the different observations of the graph signal, our kernel is not necessarily defined or dictated by the underlying graph. This is in contrast with the prior works where the kernel matrix is an explicit function of the graph adjacency matrix and the kernel is across the different nodes of the same graph.  In several applications where the input is also a graph signal, experiments with real-world datasets show that our method performs better than those which exploit the graph structure in the input.} 
%{\color{black}In experiments where both the input and output are nodes of the same graph carrying the same physical quantities, our approach performs reasonably well even when the relative number of nodes observed to the total number of nodes is $50\%$ or lesser.}

{\color{black} The success of our method can be attributed to a standard machine learning concept where a set of training data is used to learn a regression model and then the trained model is used to make predictions on the test data. The prediction at each test datapoint is made independent of the other test datapoints, and is based only on the kernels between the training datapoints and the relevant test datapoint (and not all the test datapoints together). In contrast, in the prior works involving kernels and graph signals \cite{kergraph1,kergraph4}, the estimation of the graph signal value at even one of the nodes involves the computation of the entire kernel matrix for all the nodes over the graph, and not just over the input or the training nodes. In other words, they employ the entire kernel matrix across all the available training and test datapoints and do not treat a relevant test datapoint independently with respect to the other test datapoints. This is true even for the cases where one predicts a subset of the unobserved nodes or test datapoints.}

{\color{black} Further, the independent treatment of the test datapoints allows us to use our regression method on any number of test datapoints. The method does not assume that the number of test datapoints is known from the beginning. Therefore, our method scales well with a large amount of test datapoints and naturally extends to a dynamic tracking setup, such as the Gaussian process model \cite{Arun_GPG}. On the other hand, the works of \cite{kergraph1,kergraph2,kergraph3,kergraph4} assume that the number of test datapoints is specified from the beginning. 
}

	\subsection{Signal processing over graphs}
	We next briefly review some of the basic concepts from graph signal processing.
	Let $G=(\mathcal{V},\mathcal{E},\mathbf{A})$ denote a graph with $M$ nodes indexed by the vertex set $\mathcal{V}=\{1,\cdots, M\}$.  Let $\mathcal{E}$ and $\mathbf{A}$ denote the edge set containing pairs of nodes, and the weighted adjacency matrix, respectively. The $(i,j)$th entry of the adjacency matrix $\textbf{A}(i,j)$ denotes the strength of the edge between the $i$th and $j$th nodes. {\color{black}There exists an edge between the $i$th and the $j$th nodes if $\mathbf{A}(i,j)>0$ and the edge pair $(i,j)\in\mathcal{E} \iff \mathbf{A}(i,j)\neq 0$.} In our analysis, we consider only undirected graphs with symmetric edge-weights or $\mathbf{A}=\mathbf{A}^\top$. The graph-Laplacian matrix $\mathbf{L}$ of the graph $G$ is then defined as
	\begin{equation}
	\mathbf{L=D-A},\nonumber
	\end{equation}
	where $\mathbf{D}$ is the diagonal degree matrix with the $i$th diagonal element given by the sum of the elements in the $i$th row of $\mathbf{A}$.  
	A vector $\mathbf{x}=[x(1)\,x(2)\,\cdots x(M)]^\top\in\mathbb{R}^M$ is said to be a graph signal if  {\color{black} $x(i)$ denotes the value of the signal at the $i$th node of $\mathcal{G}$.} The quadratic form of $\mathbf{x}$ with $\mathbf{L}$ is given by 
	\begin{equation}
	\mathbf{x}^\top\mathbf{L}\mathbf{x}=\sum_{(i,j)\in\mathcal{E}} \mathbf{A}(i,j)(x(i)-x(j))^2.\nonumber
	\end{equation}
	We observe that $\mathbf{x}^\top\mathbf{L}\mathbf{x}$ is minimized when the signal $\mathbf{x}$ takes the same value across all the connected nodes, which agrees with the intuitive concept of a smooth signal. In general, a graph signal is said to be smooth or a low-frequency signal if it has similar values across all the connected nodes in a graph, and is said to be a high-frequency signal if it has dissimilar values across the connected nodes, $\mathbf{x}^\top\mathbf{L}\mathbf{x}$ being the measure of similarity.
	% In other words, $\mathbf{x}$ is low-frequency signal if $\xi$ is small and a high-frequency signal if $\xi$ is large. 
	This motivates the use of $\mathbf{x}^\top\mathbf{L}\mathbf{x}$ as a constraint in the applications where  either the signal $\mathbf{x}$ or the graph-Laplacian $\mathbf{L}$ is to be estimated\cite{Dong:2016,graphlearn2}. The eigenvectors of $\mathbf{L}$ are referred to as the graph Fourier transform basis for $G$, and the corresponding eigenvalues are referred to as the graph frequencies. The smaller eigenvalues (the smallest being zero by construction) are referred to as the low frequencies since the corresponding eigenvectors {\color{black} result in small values of the quadratic form of $\mathbf{L}$,} and vary smoothly over the nodes. Similarly, the larger eigenvalues are referred to as the high frequencies. Then, a smooth graph signal is one which has the energy of the GFT coefficients predominantly in the low graph frequencies.

	\section{Kernel Regression over Graphs}
	\label{sec:Kernel_Regression_over_Graphs}
	
	\subsection{{\color{black}Linear basis model for regression} over graphs}
	\label{GLRsec}
	Let $\{\mathbf{x}_n\}_{n=1}^N$ denote a set of $N$ input observations. Each input $\mathbf{x}_n$ is paired with a target $\mathbf{t}_n\in\mathbb{R}^M$. {\color{black} Our goal is to model the target $\mathbf{t}_n$ with $\mathbf{y}_n$ given by}:
	%We model the target $\mathbf{t}_n$ using linear regression model
	%\begin{equation}
	%\mathbf{t}_n=\mathbf{y}(\mathbf{x}_n,\mathbf{W}) + \mathbf{e}_n,
	%\end{equation}
	%where $\mathbf{e}_n$ denotes additive noise which follows the isotropic Gaussian distribution $\mathcal{N}(\mathbf{0},\sigma^2\mathbf{I}_M)$ such that 
	\begin{equation}
	\mathbf{y}_n %\triangleq\mathbf{y}(\mathbf{x}_n,\mathbf{W})
	=\mathbf{W}^\top\pmb\phi(\mathbf{x}_n),
	\label{eq:regression_output}
	\end{equation}
	\begin{figure*}[t]
		\begin{eqnarray}
		\label{eq:cost_function_w}
		\begin{array}{rcl}
		C(\mathbf{W}) & = & \displaystyle\sum_n \|\mathbf{t}_n-\mathbf{W}^\top\pmb\phi(\mathbf{x}_n)\|_2^2+\alpha \,\mbox{tr}(\mathbf{W}^\top\mathbf{W}) + \beta \,\sum_n \pmb\phi(\mathbf{x}_n)^\top \mathbf{W L}\mathbf{W}^\top \pmb\phi(\mathbf{x}_n) \\
		%=& \sum_n \|\mathbf{t}_n\|_2^2-2\sum_n \pmb\phi(\mathbf{x}_n)^T\mathbf{W}\mathbf{t}_n\nonumber\\
		%&+\sum_n  \pmb\phi(\mathbf{x}_n)^T\mathbf{W}\mathbf{W}^\top\pmb\phi(\mathbf{x}_n) +\alpha\, \mbox{tr}(\mathbf{W}^\top\mathbf{W}) \nonumber\\
		%&+  \beta \,\sum_n \pmb\phi(\mathbf{x}_n)^T \mathbf{W L}\mathbf{W}^\top \pmb\phi(\mathbf{x}_n)\nonumber\\
		& = & \displaystyle\sum_n \|\mathbf{t}_n\|_2^2-2 \, \mbox{tr}\left( \sum_n \pmb\phi(\mathbf{x}_n)^\top\mathbf{W}\mathbf{t}_n\right) 
		+\mbox{tr}\left( \displaystyle\sum_n  \pmb\phi(\mathbf{x}_n)^\top\mathbf{W}\mathbf{W}^\top\pmb\phi(\mathbf{x}_n) \right) +\alpha \,\mbox{tr}(\mathbf{W}^\top\mathbf{W}) \\
		& & +  \beta \,\mbox{tr}\left( \displaystyle\sum_n \pmb\phi(\mathbf{x}_n)^\top \mathbf{W L}\mathbf{W}^\top \pmb\phi(\mathbf{x}_n)\right) \\
		& = & \displaystyle\sum_n \|\mathbf{t}_n\|_2^2-2 \, \mbox{tr}\left(\sum_n\mathbf{t}_n\pmb\phi(\mathbf{x}_n)^\top \mathbf{W} \right) + \mbox{tr}\left(  \mathbf{W}\mathbf{W}^\top\sum_n \pmb\phi(\mathbf{x}_n) \pmb\phi(\mathbf{x}_n)^\top\right) +\alpha\, \mbox{tr}(\mathbf{W}^\top\mathbf{W}) \\
		& & +  \beta\, \mbox{tr}\left(  \mathbf{W L}\mathbf{W}^\top \displaystyle\sum_n  \pmb\phi(\mathbf{x}_n)\pmb\phi(\mathbf{x}_n)^\top\right) \\
		& = & \displaystyle\sum_n \|\mathbf{t}_n\|_2^2-2\,\mbox{tr}\left(\mathbf{T}^\top\mathbf{\Phi}\mathbf{W} \right) +\mbox{tr}\large(  \mathbf{W}^\top\mathbf{\Phi}^\top\mathbf{\Phi}\mathbf{W}\large) + \alpha\, \mbox{tr}(\mathbf{W}^\top\mathbf{W}) +  \beta\, \mbox{tr}\left(  \mathbf{W}^\top\mathbf{\Phi}^\top\mathbf{\Phi}\mathbf{W}\mathbf{ L}\right).
		% \nonumber\\
		\end{array}
		\end{eqnarray}
	\end{figure*}
	where $\pmb\phi(\mathbf{x}_n)\in\mathbb{R}^K$ is a known function of $\mathbf{x}_n$ and $\mathbf{W}\in\mathbb{R}^{K\times M}$ denotes the regression coefficient matrix. {Equation \eqref{eq:regression_output} is referred to as the \color{black} linear basis model for regression, often shortened to linear regression {\color{black}in the machine learning parlance }(cf. Chapters 3 and 6 in \cite{Bishop}). For brevity, we hereafter follow this shortened nomenclature and refer to the outcome of using \eqref{eq:regression_output} as linear regression.} Our central assumption is that the target $\mathbf{y}_n$ is a smooth signal over an underlying graph $G$ with $M$ nodes. We learn the optimal parameter matrix $\mathbf{W}$ by minimizing the following cost function with respect to $\mathbf{W}$:
	\begin{equation}
	\label{eq:cost}
	C(\mathbf{W})=\sum_{n=1}^N \|\mathbf{t}_n-\mathbf{y}_n\|_2^2 + \alpha \mbox{tr}(\mathbf{W}^\top\mathbf{W})+\beta \sum_{n=1}^N \mathbf{y}_n^\top\mathbf{L}\mathbf{y}_n,
	\end{equation}
	where the regularization coefficients $\alpha,\beta\geq 0$, $\mbox{tr}(\cdot)$ denotes the trace operator, and $\|\mathbf{x}\|_2$ denotes the $\ell_2$ norm of $\mathbf{x}$, {\color{black} and we emphasize that $\mathbf{y}_n$ is a function of $\mathbf{W}$}. {\color{black}The cost in \eqref{eq:cost} is convex in $\mathbf{W}$, since $\mathbf{L}$ is positive semidefinite on the virtue of it being the graph-Laplacian matrix}. The choice of $\alpha,\beta$ depends on the problem, and in our analysis we compute these parameters through {\color{black} crossvalidation}. The penalty $\mbox{tr}(\mathbf{W}^\top\mathbf{W}) = \|\mathbf{W}\|_F^2$ ensures that $\mathbf{W}$ remains bounded. The penalty or regularization $ \mathbf{y}_n^\top\mathbf{L}\mathbf{y}_n$ enforces $\mathbf{y}_n$ to be smooth over $G$. {\color{black} We note that the smoothness over a graph could be quantified in a number of alternative ways, specially if domain-specific knowledge is available. However, since the graph-Laplacian based regularization has been the most popular metric in the graph signal processing literature with respect to undirected graphs, we employ the same in our work. Another aspect is that $ \mathbf{y}_n^\top\mathbf{L}\mathbf{y}_n$ being quadratic in $\mathbf{y}_n$ helps us arrive at a unique and closed-form solution for the regression model, as we show next.}
	We define matrices $\mathbf{T}$, $\mathbf{Y}$ and $\mathbf{\Phi}$ as follows:
	\begin{eqnarray}
	\begin{array}{rcl}
	\mathbf{T}&=&[\mathbf{t}_1\,\mathbf{t}_2\,\cdots \mathbf{t}_N]^{\top} \in\mathbb{R}^{N \times M}, \\
	\mathbf{Y}&=&[\mathbf{y}_1\,\mathbf{y}_2\,\cdots \mathbf{y}_N]^{\top} \in\mathbb{R}^{N \times M},\\
	\mathbf{\Phi}&=&[\pmb\phi(\mathbf{x}_1)\,\,\pmb\phi(\mathbf{x}_2)\,\,\cdots \, \pmb\phi(\mathbf{x}_N)]^{\top} \in\mathbb{R}^{N \times K}.
	\end{array}
	\label{eq:matrix_definitions}
	\end{eqnarray}
	Using \eqref{eq:regression_output} and \eqref{eq:matrix_definitions}, the cost function \eqref{eq:cost} is expressible as \eqref{eq:cost_function_w} where we use the properties of the matrix trace operation.
Since the cost function is quadratic {\color{black} and convex} in $\mathbf{W}$, we get the optimal and unique solution by setting the gradient of ${C}$ with respect to $\mathbf{W}$ equal to zero.
	Using the following matrix derivative relations
	\begin{eqnarray*}
		\begin{array}{rcl}
			\displaystyle\frac{\partial}{\partial \mathbf{W}} \mbox{tr}\left( \mathbf{M}_1 \mathbf{W} \right) & = & \mathbf{M}_1^{\top}, \\
			\displaystyle\frac{\partial}{\partial \mathbf{W}} \mbox{tr}\left( \mathbf{W}^{\top} \mathbf{M}_1 \mathbf{W} \mathbf{M}_2 \right) & = &  \mathbf{M}_1^{\top} \mathbf{W} \mathbf{M}_2^{\top} + \mathbf{M}_1 \mathbf{W} \mathbf{M}_2,
		\end{array}
	\end{eqnarray*}
	where $\mathbf{M}_1$ and $\mathbf{M}_2$ are matrices, and setting $\frac{\partial C}{\partial \mathbf{W}} =0$ we get that
	\begin{eqnarray}
	\begin{array}{l}
	-\mathbf{\Phi}^\top\mathbf{T}+\mathbf{\Phi}^\top\mathbf{\Phi W} +\alpha \mathbf{W} +\beta \mathbf{\Phi}^\top\mathbf{\Phi W L}= \mathbf{0}, \\
	\mathrm{or,} \,\,\,\, \left(\mathbf{\Phi}^\top\mathbf{\Phi}+\alpha \mathbf{I}_K\right)\mathbf{W}+\beta \mathbf{\Phi}^\top\mathbf{\Phi W L}=\mathbf{\Phi}^\top \mathbf{T}. \\    	 \end{array}
	\label{eq:J_ww}
	\end{eqnarray}
	On vectorizing both sides of \eqref{eq:J_ww}, we get that
	\begin{align}
	\mbox{vec}(\mathbf{\Phi}^\top\mathbf{T}) \! = \! \left[\!(\mathbf{I}_M \! \otimes \! (\!\mathbf{\Phi}^\top\mathbf{\Phi} \!+ \! \alpha\mathbf{I}_K) \!) \! + \! (\beta \mathbf{L} \! \otimes \! \mathbf{\Phi}^\top\mathbf{\Phi})\!\right] \! \mbox{vec}(\mathbf{W}), \nonumber 
	\end{align}
	where $\mbox{vec}(\cdot)$ denotes the standard vectorization operator and $\otimes$ denotes the Kronecker product operation \cite{Loan1}. Then, the optimal $\mathbf{W}$, denoted by $\overset{\star}{\mathbf{W}}$, follows the relation:
	\begin{equation}
	\mbox{vec}(\overset{\star}{\mathbf{W}}) \! = \!  \left[\!(\mathbf{I}_M \! \otimes \! (\!\mathbf{\Phi}^\top\mathbf{\Phi} \!+ \! \alpha\mathbf{I}_K) \!) \! + \! (\beta \mathbf{L} \! \otimes \! \mathbf{\Phi}^\top\mathbf{\Phi})\!\right]^{-1}   \mbox{vec}(\mathbf{\Phi}^\top\mathbf{T}).\quad\nonumber
	\end{equation}
	The predicted target for a new input $\mathbf{x}$ is then given by
	\begin{equation}
	\label{eq: lrg}
	\overset{\star}{\mathbf{t}}= \overset{\star}{\mathbf{W}}^{\top} \pmb\phi(\mathbf{x}).
	\end{equation} 
From \eqref{eq: lrg}, it appears that the proposed target prediction approach requires the explicit knowledge of the function $\pmb\phi(\cdot)$. We next show that using the `kernel trick' or `kernel substitution' this explicit requirement of $\pmb\phi(\cdot)$ is circumvented and that the target prediction may be done using only the knowledge of the inner products $\pmb\phi(\mathbf{x}_m)^\top\pmb\phi(\mathbf{x}_n)$, $\forall m,n$. Towards this end, we next discuss a dual representation of the cost in \eqref{eq:cost}. We hereafter refer to \eqref{eq: lrg} as the output of the linear regression over graphs (LRG).
	
	\subsection{Dual representation of cost using kernel trick}
	
We now use the substitution $\mathbf{W}=\mathbf{\Phi}^\top \mathbf{\Psi}$ and express the cost function in terms of the parameter $\mathbf{\Psi}\in\mathbb{R}^{N\times M }$. 
 This substitution is motivated by observing that on rearranging the terms in \eqref{eq:J_ww}, we get that
	\begin{equation}
	\mathbf{W} = \mathbf{\Phi}^\top\pmb\Psi,\nonumber
	\end{equation}
	where $ \pmb\Psi=\left[ \frac{1}{\alpha} \left( \mathbf{T} - \beta \mathbf{\Phi W L}   - \mathbf{\Phi} \right) \right]$. On substituting $\mathbf{W}=\mathbf{\Phi}^\top \mathbf{\Psi}$ in \eqref{eq:cost_function_w}, where $\pmb\Psi$ becomes the dual parameter matrix that we wish to learn, and omitting terms that do not depend on $\mathbf{W}$, we get that
	\begin{align}
		\label{J_dual}
	C(\mathbf{\Psi})=&-2\mbox{tr}\left(\mathbf{T} ^T\mathbf{\Phi}\mathbf{\Phi}^\top\mathbf{\Psi}\right) +\mbox{tr}\left( \mathbf{\Psi}^\top\bf{\Phi\Phi^\top}\bf{\Phi\Phi^\top}\mathbf{\Psi}\right) \nonumber\\
	& +\alpha \,\mbox{tr}(\mathbf{\Psi}^\top\bf{\Phi\Phi^\top}\mathbf{\Psi}) +  \beta\, \mbox{tr}\left( \mathbf{\Psi}^\top\bf{\Phi\Phi^\top}\bf{\Phi\Phi^\top}\mathbf{\Psi}\mathbf{ L}\right)\nonumber\\
	=&-2\mbox{tr}\left(\mathbf{T}^\top\bf{K}\mathbf{\Psi} \right) +\mbox{tr}\left( \mathbf{\Psi}^\top\bf{K}\bf{K}\mathbf{\Psi}\right) \nonumber\\
	& +\alpha\, \mbox{tr}(\mathbf{\Psi}^\top\bf{K}\mathbf{\Psi}) +  \beta \,\mbox{tr}\left( \mathbf{\Psi}^\top\bf{K}\bf{K}\mathbf{\Psi}\mathbf{ L}\right),
	% \nonumber\\
	\end{align}
	where $\mathbf{K}=\mathbf{\Phi}\mathbf{\Phi}^\top\in\mathbb{R}^{N\times N}$ denotes the kernel matrix for the training samples such that its $(m,n)$th entry is given by
	\begin{equation}
	k_{m,n}=\pmb\phi(\mathbf{x}_m)^\top\pmb\phi(\mathbf{x}_n).\nonumber
	\end{equation}
	{\color{black} Equation \eqref{J_dual} is referred to as a dual representation of \eqref{eq:cost_function_w} in the kernel regression literature (cf. Chapter 6 of \cite{Bishop}).}
	Taking the derivative of $C(\mathbf{\Psi})$ with respect to $\mathbf{\Psi}$ and setting it to zero, we get that
	\begin{eqnarray}
	%\label{a_1}
	(\mathbf{I}_M\otimes (\mathbf{K}+\alpha\mathbf{I}_N))\mbox{vec}(\mathbf{\Psi})+(\beta \mathbf{L}\otimes \mathbf{K})\mbox{vec}(\mathbf{\Psi})=\mbox{vec}(\mathbf{T}), \,\, \mbox{or}\nonumber\\
	\left[(\mathbf{I}_M\otimes (\mathbf{K}+\alpha\mathbf{I}_N))+(\beta \mathbf{L}\otimes \mathbf{K})\right]\mbox{vec}(\mathbf{\Psi})=\mbox{vec}(\mathbf{T}), \quad\mbox{or}\nonumber\\
	\mbox{vec}(\mathbf{\Psi})=\left[(\mathbf{I}_M\otimes (\mathbf{K}+\alpha\mathbf{I}_N))+(\beta \mathbf{L}\otimes \mathbf{K})\right]^{-1}\mbox{vec}(\mathbf{T}).\nonumber
	\end{eqnarray}
	We define the matrices
	\begin{equation}
	\begin{array}{rcl}
	\mathbf{B} & = & (\mathbf{I}_M\otimes (\mathbf{K}+\alpha\mathbf{I}_N)), \\
	\mathbf{C} & = & (\beta \mathbf{L}\otimes \mathbf{K}).  
	\end{array}
	\label{eq:B_and_C}
	\end{equation}	
	Then, we have that
	\begin{align}
	\mbox{vec}(\mathbf{\Psi})=\left(\mathbf{B+C}\right)^{-1}\mbox{vec}(\mathbf{T}).
	\label{eq:vec_psi}
	\end{align}
	Once $\pmb\Psi$ is computed, the predicted output of the kernel regression for a new test input $\mathbf{x}$ is given by
	\begin{align}
	\begin{array}{rcl}
	\mathbf{y}&=&\mathbf{W}^{\top} \, \pmb\phi(\mathbf{x}) =\mathbf{\Psi}^{\top} \, \mathbf{\Phi}\pmb\phi(\mathbf{x}) \\
	&=& \mathbf{\Psi}^{\top}\mathbf{k}(\mathbf{x}) \\
	& = & \left(\mbox{mat} \left( \left(\mathbf{B+C}\right)^{-1}\mbox{vec}(\mathbf{T}) \right) \right)^{\top}\mathbf{k}(\mathbf{x}),
	\end{array}
	\label{eq:KRG_output}
	\end{align}
	where $\mathbf{k}(\mathbf{x})=[k_1(\mathbf{x}), k_2(\mathbf{x}),\cdots, k_N(\mathbf{x})]^\top$ and $k_n(\mathbf{x})=\pmb\phi(\mathbf{x}_n)^\top\pmb\phi(\mathbf{x})$. Here $\mbox{mat}(\cdot)$ denotes the reshaping operation of an argument vector into an appropriate matrix of size $N \times M$ by concatenating the subsequent $N$ length sections as the columns. We refer to \eqref{eq:KRG_output} as the output of the method named  \textit{kernel regression over graphs} (KRG). The kernel regression is arrived at by noting that the entire formulation remains valid if the inner products $\pmb\phi(\mathbf{x}_m)^\top\pmb\phi(\mathbf{x}_n)$ are {\color{black}replaced} with a general kernel function associating pairs of the inputs $\mathbf{x}_m$ and $\mathbf{x}_n$.
	
	{\color{black} In general, a variety of valid kernel functions may be employed. Any kernel function $k(\mathbf{x},\mathbf{x}')$ is a valid kernel as long as it can be expressed in the form $k(\mathbf{x},\mathbf{x}')=\pmb\phi(\mathbf{x})^\top\pmb\phi(\mathbf{x}')$, and the associated kernel matrix $\mathbf{K}$ is positive semi-definite for all observation sizes \cite{Bishop}. The Gaussian kernel is a popularly employed kernel and we use the same in our experiments later.}  {\color{black}We note that KRG is a generalization over the conventional kernel regression (KR), where the latter does not use any knowledge of the underlying graph structure}. On setting $\beta=0$, the KRG output \eqref{eq:KRG_output} reduces to the conventional KR output as follows 
	\begin{eqnarray}
	%\begin{array}{rcl}
	\mathbf{y} & = & \left(\mbox{mat} \left( \left(\mathbf{B+C}\right)^{-1}\mbox{vec}(\mathbf{T}) \right) \right)^{\top}\mathbf{k}(\mathbf{x})\nonumber \\
	& = & \left(\mbox{mat} \left( \left(\mathbf{B}\right)^{-1}\mbox{vec}(\mathbf{T}) \right) \right)^{\top}\mathbf{k}(\mathbf{x}) \nonumber\\
	& = & \left(\mbox{mat} \left( \left(\mathbf{I}_M\otimes (\mathbf{K}+\alpha\mathbf{I}_N) \right)^{-1}\mbox{vec}(\mathbf{T}) \right) \right)^{\top}\mathbf{k}(\mathbf{x})\nonumber \\
	& = & \left(\mbox{mat} \left( ( \mathbf{I}_M^{-1} \otimes (\mathbf{K}+\alpha\mathbf{I}_N)^{-1} ) \mbox{vec}(\mathbf{T}) \right) \right)^{\top}\mathbf{k}(\mathbf{x})\nonumber \\
	& = &  \left( (\mathbf{K}+\alpha\mathbf{I}_N)^{-1} \mathbf{T} \right)^{\top} \mathbf{k}(\mathbf{x})\nonumber \\
	& = & \mathbf{T}^{\top} (\mathbf{K}+\alpha\mathbf{I}_N)^{-1} \mathbf{k}(\mathbf{x}),
	%	\end{array}
	\label{eq:KR_output}
	\end{eqnarray}
	where we have used the Kronecker product equality: $\mbox{vec}((\mathbf{K}+\alpha\mathbf{I}_N)^{-1} \mathbf{T}) = ( \mathbf{I}_M^{-1} \otimes (\mathbf{K}+\alpha\mathbf{I}_N)^{-1} ) \mbox{vec}(\mathbf{T})$.
	Further, we note that KRG reduces to LRG on setting $\mathbf{K}=\mathbf{ \Phi\Phi}^\top$.
	
	\subsection{Interpretation of KRG -- a smoothing effect}
	\label{smoothing}
	
	We next discuss how the output of KRG is smooth across the training samples $\{1,\cdots,N\}$ and over the $M$ nodes of the graph.  Before proceeding with KRG, we review a similar property exhibited by the KR output \cite{RasmussenGP}. Using \eqref{eq:KR_output} and concatenating the KR outputs for the $N$ training samples, we get that
	\begin{eqnarray*}
		%\begin{array}{l}
		[ \mathbf{y}_1 \, \mathbf{y}_2 \, \hdots \, \mathbf{y}_N ] = \mathbf{T}^{\top} (\mathbf{K}+\alpha\mathbf{I}_N)^{-1} [ \mathbf{k}(\mathbf{x}_1) \, \mathbf{k}(\mathbf{x}_2) \, \hdots \mathbf{k}(\mathbf{x}_N) ] \qquad\\
		\mbox{or}, \,\,\,\, \mathbf{Y} = \mathbf{K} (\mathbf{K}+\alpha\mathbf{I}_N)^{-1} \mathbf{T},\qquad\qquad
		%\end{array}
	\end{eqnarray*}
	where we use $\mathbf{K} =  [ \mathbf{k}(\mathbf{x}_1) \, \mathbf{k}(\mathbf{x}_2) \, \hdots \mathbf{k}(\mathbf{x}_N) ]$.
	Assuming $\mathbf{K}$ is diagonalizable, let $\mathbf{K}=\mathbf{U}\mathbf{J}_K\mathbf{U}^{\top}$, where $\mathbf{J}_K$ and $\mathbf{U}$ denote the eigenvalue and the eigenvector matrices of $\mathbf{K}$, respectively. {\color{black}Let $\tilde{\mathbf{y}}(m)=[y_1(m),y_2(m),\cdots,y_N(m)]^\top$ $(m\in\{1,\cdots,M\})$ denote the $m$th column of $\mathbf{Y}$. We note that $\tilde{\mathbf{y}}(m)$ consists of values of the $m$th component of $\mathbf{y}$ collected over all the time instances. Then, we have that
	\begin{equation}
\tilde{\mathbf{y}}(m)=\sum_{i=1}^{N}\frac{\theta_i}{\theta_i+\alpha}[\mathbf{u}_i^\top\tilde{\mathbf{t}}(m)]\mathbf{u}_i,\nonumber
	\end{equation}
	where $\theta_i$ and $\mathbf{u}_i$ denote the $i$th eigenvalue and eigenvector of $\mathbf{K}$, respectively, and $\tilde{\mathbf{t}}(m)$ denotes the vector containing the $m$th component of the target vector for all the training samples ($m$th column of $\mathbf{T}$). Thus, we observe that the KR output performs a shrinkage of $\tilde{\mathbf{t}}(m)$ along the various eigenvector directions $\mathbf{u}_i$ for each $m$. The contribution from the eigenvectors corresponding to the smaller eigenvalues $\theta_i< \alpha$ are effectively {\color{black}eliminated}, and only those corresponding to the larger eigenvalues $\theta_i>\alpha$ are retained.} Since the eigenvectors corresponding to the larger eigenvalues of $\mathbf{K}$ represent smooth variations across the observations $\{1,\cdots,N\}$, we observe that KR performs a smoothing of $\mathbf{t}_n$. We next show that such is also the case with KRG: KRG acts as a smoothing filter across both the observations and the graph nodes. 
	Using \eqref{eq:KRG_output} and concatenating the KRG outputs, we have that
	\begin{eqnarray}
	\label{eq:krg_op}
	[ \mathbf{y}_1 \, \mathbf{y}_2 \, \hdots \, \mathbf{y}_N ] &=& \mathbf{\Psi}^{\top} [ \mathbf{k}(\mathbf{x}_1) \, \mathbf{k}(\mathbf{x}_2) \, \hdots \mathbf{k}(\mathbf{x}_N) ] \\
	\mbox{or}, \,\,\,\, \mathbf{Y} &=& \mathbf{K} \mathbf{\Psi}. \nonumber
	\end{eqnarray}
	On vectorizing both sides of (\ref{eq:krg_op}), we get that
	\begin{eqnarray}
	\begin{array}{rl} 
	\mbox{vec}(\mathbf{Y}) & = (\mathbf{I}_M\otimes \mathbf{K})\mbox{vec}(\mathbf{\Psi}) \\
	& \overset{(a)}{=} (\mathbf{I}_M\otimes \mathbf{K}) \left(\mathbf{B+C}\right)^{-1}\mbox{vec}(\mathbf{T}),
	\end{array}
	\label{eq:vec_Y}	
	\end{eqnarray}
	where we have used \eqref{eq:vec_psi} in (a). Let $\mathbf{L}$ be diagonalizable with the eigendecomposition: 
	\begin{eqnarray}
	\mathbf{L}=\mathbf{V}\mathbf{J}_L\mathbf{V}^{\top},\,\,\,\,\nonumber
	%\mathbf{K}=\mathbf{U}\mathbf{J}_K\mathbf{U}^{T},\nonumber
	\end{eqnarray}
	where $\mathbf{J}_L$ and $\mathbf{V}$ denote the eigenvalue and the eigenvector matrices of $\mathbf{L}$, respectively. We also assume that $\mathbf{K}$ is diagonalizable as earlier. Let $\lambda_i$ and $\mathbf{v}_i$ denote the $i$th eigenvalue and the $i$th eigenvector of $\mathbf{L}$, respectively. Then, we have that
	\begin{eqnarray}
	\mathbf{V}&=&[\mathbf{v}_1\,\mathbf{v}_2\cdots \mathbf{v}_N]\,\, \mbox{and}\,\, \mathbf{J}_L=\mbox{diag}(\lambda_1,\lambda_2,\cdots,\lambda_N),\nonumber\\ 
	\mathbf{U}&=&[\mathbf{u}_1\,\mathbf{u}_2\cdots \mathbf{u}_M]\,\, \mbox{and}\,\, \mathbf{J}_K=\mbox{diag}(\theta_1,\theta_2,\cdots,\theta_M).\qquad \nonumber
	\end{eqnarray}
	Now, using \eqref{eq:B_and_C}, we have
	\begin{eqnarray}
	\label{eq:krg_smooth}
	\mathbf{B} + \mathbf{C}  & =&  (\mathbf{I}_M \otimes (\mathbf{K}+\alpha\mathbf{I}_N)) + (\beta \mathbf{L}\otimes \mathbf{K}) \nonumber\\
	& =&  [(\mathbf{V} \mathbf{I}_M  \mathbf{V}^{\top}) \otimes (\mathbf{U} (\mathbf{J}_K +\alpha\mathbf{I}_N) \mathbf{U}^{\top})] \nonumber \\
	&& \hspace{0.5cm} + [\beta(\mathbf{V}\mathbf{J}_L\mathbf{V}^{\top}) \otimes (\mathbf{U}\mathbf{J}_K\mathbf{U}^{\top})] \nonumber\\
	& \overset{(a)}{=}&  [(\mathbf{V} \otimes \mathbf{U}) (\mathbf{I}_M \otimes (\mathbf{J}_K +\alpha\mathbf{I}_N)) (\mathbf{V}^{\top} \otimes \mathbf{U}^{\top})]\nonumber \\
	&& \hspace{0.5cm} + [\beta (\mathbf{V} \otimes \mathbf{U}) (\mathbf{J}_L \otimes \mathbf{J}_K) (\mathbf{V}^{\top} \otimes \mathbf{U}^{\top})] \nonumber\\
	& =& \mathbf{Z} \mathbf{J} \mathbf{Z}^{\top},
	\end{eqnarray}
	where $\mathbf{Z} = \mathbf{V} \otimes \mathbf{U}$ is the eigenvector matrix and $\mathbf{J}$ is the diagonal eigenvalue matrix given by
	\begin{eqnarray}
	\label{eq:Jexp1}
	\mathbf{J} = (\mathbf{I}_M \otimes (\mathbf{J}_K +\alpha\mathbf{I}_N)) + \beta (\mathbf{J}_L \otimes \mathbf{J}_K).
	\end{eqnarray}
	In (\ref{eq:krg_smooth})(a), we have used the distributivity property of the Kronecker product: $(\mathbf{M}_1 \otimes \mathbf{M}_2) (\mathbf{M}_3 \otimes \mathbf{M}_4) = \mathbf{M}_1\mathbf{M}_3 \otimes \mathbf{M}_2 \mathbf{M}_4$ where $\{\mathbf{M}_i\}_{i=1}^{4}$ are four matrices. We note that $\mathbf{J}$ is a diagonal matrix of size $MN$. Let $\mathbf{J}=\mbox{diag}(\eta_1 , \eta_2, \hdots, \eta_{MN})$. Then, $\eta_i$ is a function of $\{\lambda_j\},\{\theta_j\},\alpha,\beta$. On dropping the subscripts for simplicity, we observe that any eigenvalue $\eta_i$ has the form
	\begin{eqnarray}
	\eta = (\theta+\alpha) + \beta(\lambda\theta).\nonumber
	\end{eqnarray}
	where $\theta$ and $\lambda$ are the appropriate eigenvalues of $\mathbf{K}$ and $\mathbf{L}$, respectively.
	Similarly, we have that
	\begin{eqnarray}
	\label{eq:Jexp2}
	\begin{array}{rcl}
	(\mathbf{I}_M \otimes \mathbf{K}) & = & (\mathbf{V} \mathbf{I}_M  \mathbf{V}^{\top}) \otimes (\mathbf{U} \mathbf{J}_K \mathbf{U}^{\top}) \\
	& = &  (\mathbf{V} \otimes \mathbf{U})   (\mathbf{I}_M \otimes \mathbf{J}_K) (\mathbf{V}^{\top} \otimes \mathbf{U}^{\top}) \\
	& = & \mathbf{Z} (\mathbf{I}_M \otimes \mathbf{J}_K) \mathbf{Z}^{\top},
	\end{array}
	\end{eqnarray}
	and note that $(\mathbf{I}_M \otimes \mathbf{J}_K)$ is also a diagonal matrix of size $MN$.
	Then, on substituting (\ref{eq:Jexp1}) and (\ref{eq:Jexp2}) in \eqref{eq:vec_Y}, we get that
	\begin{eqnarray}
	\begin{array}{rl} 
	\mbox{vec}(\mathbf{Y}) & = (\mathbf{I}_M\otimes \mathbf{K}) \left(\mathbf{B+C}\right)^{-1}\mbox{vec}(\mathbf{T}) \\
	& = (\mathbf{Z} (\mathbf{I}_M \otimes \mathbf{J}_K) \mathbf{Z}^{\top}) (\mathbf{Z} \mathbf{J}^{-1} \mathbf{Z}^{\top})  \mbox{vec}(\mathbf{T}) \\
	& =  (\mathbf{Z} (\mathbf{I}_M \otimes \mathbf{J}_K)  \mathbf{J}^{-1} \mathbf{Z}^{\top})  \mbox{vec}(\mathbf{T}).
	\end{array}
	\label{eq:vec_Y_simplified}	
	\end{eqnarray}
	We note again that $(\mathbf{I}_M \otimes \mathbf{J}_K)  \mathbf{J}^{-1}$ is a diagonal matrix with size $MN$. Let $(\mathbf{I}_M \otimes \mathbf{J}_K)  \mathbf{J}^{-1}=\mbox{diag}(\zeta_1 , \zeta_2, \hdots, \zeta_{MN})$. Then, on dropping the subscripts, any $\eta_i$ is of the form
	\begin{eqnarray}
	\zeta = \frac{\theta}{\eta} = \frac{\theta}{(\theta+\alpha) + \beta(\lambda\theta)}.\nonumber
	\end{eqnarray}
	From \eqref{eq:vec_Y_simplified}, we have that 
	\begin{eqnarray}
	\label{vec_a}
	\mbox{vec}\left(\mathbf{Y}\right)\nonumber=\sum_{i=1}^{MN}\zeta_i\mathbf{z}_i\mathbf{z}_i^\top\mbox{vec}\left(\mathbf{T}\right),
	\end{eqnarray}
	where $\mathbf{z}_i$ are the column vectors of $\mathbf{Z}$. 
	In the case when $\zeta_i\ll1$, the component in $\mbox{vec}(\mathbf{T})$ along $\mathbf{z}_i$ is effectively eliminated. For most covariance or kernel functions $k(\cdot,\cdot)$ used in practice, the eigenvectors corresponding to the largest eigenvalues of $\mathbf{K}$ are the low-frequency components across time or observations. Similarly, the eigenvectors corresponding to the smaller eigenvalues of $\mathbf{L}$ are smooth over the graph \cite{Shuman}. We observe that the condition $\zeta\ll 1$ is achieved when $\theta$ is small {\color{black}and/or} $\lambda$ is large. This condition in turn corresponds to effectively retaining only the components of $\mbox{vec}(\mathbf{T})$ which vary smoothly across the samples $\{1,\cdots,N\}$ as well as over the $M$ nodes of the graph.

	\subsection{Learning an underlying graph}
	\label{learngraph}
	In developing KRG, we have so far assumed that the underlying graph is known \textit{apriori} in terms of the graph-Laplacian matrix $\mathbf{L}$. Such an assumption may not hold true in many practical applications, since there is not necessarily one best graph to describe the given networked data. This motivates us to develop a joint learning approach where we learn both the graph-Laplacian $\mathbf{L}$ and the KRG parameter matrix $\mathbf{W}$ (or its dual representation parameter $\mathbf{\Psi}$). {\color{black}Our goal in this section is to provide a simple means of estimating a graph that helps enhance the prediction performance, if a graph is not known apriori. We note that a vast and expanding body of literature exists in the domain of estimating graphs from graph signals \cite{graphlearn2,Dong:2016,graphlearn3,graphlearn4,graphlearn5,graphlearn6,graphlearn7,graphlearn8,graphlearn9,kalofolias16}, and that many of these techniques may be used in the learning approach proposed in this section. Nevertheless, we pursue the particular approach taken in this section due to the  ease of implementation and the minimal assumptions involved. }
	
	We propose the minimization of the following cost function to achieve our goal:
	\begin{eqnarray}
	%\begin{array}{rcl}
	C(\mathbf{W},\mathbf{L}) & = & \displaystyle\sum_n \|\mathbf{t}_n-\mathbf{y}_n\|_2^2 + \alpha \, \mbox{tr}(\mathbf{W}^\top\mathbf{W}) \nonumber\\
	& & +\beta \displaystyle\sum_n \mathbf{y}_n^\top\mathbf{L}\mathbf{y}_n + \nu \, \mbox{tr}(\mathbf{L}^{\top}\mathbf{L}),\nonumber
	%\end{array}
	%\label{eq:cost_joint}
	\end{eqnarray}
	where $\nu \geq 0$.
	Since our goal is to recover an undirected graph, we impose the appropriate constraints \cite{Chung}. Firstly, any {\color{black} non-trivial graph} has a graph-Laplacian matrix $\mathbf{L}$ which is symmetric and positive semi-definite\cite{Chung}. Secondly, the vector of all ones $\mathbf{1}$ forms the eigenvector of the graph-Laplacian corresponding to the zero eigenvalue.  {\color{black} Since $\mathbf{L}=\mathbf{D}-\mathbf{A}$, we have that $\mathbf{L}$ being positive semi-definite is equivalent to the constraint that all the off-diagonal elements of $\mathbf{L}$ are non-positive. This is a simpler constraint than the direct positive semi-definiteness constraint}. Then, the solution to the joint estimation of $\mathbf{W}$ and $\mathbf{L}$ is obtained by solving the following:
	\begin{eqnarray}
	\label{eq: krg_joint_cost}
	\displaystyle\underset{\mathbf{W},\mathbf{L}}{\min} \,C(\mathbf{W},\mathbf{L}) \quad \mbox{such that} \quad {\color{black}\mathbf{L}(i,j) \leq {0}\,\,\forall i\neq j},\nonumber\\ \mathbf{L}=\mathbf{L}^{\top}, 
	\mathbf{L}\mathbf{1}=\mathbf{0}.
	\end{eqnarray}
	The optimization problem (\ref{eq: krg_joint_cost}) is jointly non-convex over $\mathbf{W}$ and $\mathbf{L}$, but {\color{black}convex on $\mathbf{W}$ given $\mathbf{L}$ and vice-versa}. Hence, we adopt an alternating minimization approach and solve (\ref{eq: krg_joint_cost}) in two steps alternatingly as follows:
	\begin{itemize}
		\item For a given $\mathbf{L}$, solve $\underset{\mathbf{W}}{\min} \,C(\mathbf{W})$ using the KRG approach of Section~\ref{sec:Kernel_Regression_over_Graphs}.
		\item Given the matrix $\mathbf{W}$, solve $\underset{\mathbf{L}}{\min} \,C(\mathbf{L})$ $ \mbox{such that} \quad {\color{black}\mathbf{L}(i,j) \leq {0}\,\,\forall i\neq j}, \mathbf{L}=\mathbf{L}^{\top}$, and $\mathbf{L}\mathbf{1}=\mathbf{0}$. Here $C(\mathbf{L}) = \beta \sum_n \mathbf{y}_n^\top\mathbf{L}\mathbf{y}_n + \nu \, \mbox{tr}(\mathbf{L}^{\top}\mathbf{L})$. 
			\end{itemize}
			We start the alternating optimization using a suitable initialization; initializing $\mathbf{L}=\mathbf{0}$ yields the KR. In order to keep the successive $\mathbf{L}$ estimates comparable, we scale $\mathbf{L}$ such that the largest eigenvalue modulus is unity at every iteration. 
          % In practice, we find the iterations to converge in five to ten iterations.

\section{Experimental results}
\label{experiments}

{\color{black}We evaluate the performance of the relevant methods under adverse conditions where we use limited training data and noisy training data.}
{\color{black}Our hypothesis is that KRG and LRG provide better prediction performance than KR and LR, respectively. A state-of-the-art method is also compared with in the experiments of graph signal reconstruction. We experiment with both synthesized and real-world signal examples.
The experiment with the synthesized data is carried out using small-world graphs; this being a standard practice in several existing works to demonstrate the efficiency of a model.
For real applications, we consider three different real-world data experiments:
\begin{enumerate}
		\item[(D1)] Prediction of the temperature as the output, using the air-pressure observations as the input, for the cities in Sweden. 
	\item[(D2)] Temperature prediction for the cities in Sweden from the current day to the next day.
%	\item[(D3)] Prediction on EEG signals from sensor data.
%	\item[(D4)] Prediction for atmospheric tracer diffusion data.
	\item[(D3)] Prediction for the fMRI voxel intensities of the cerebellum region.
\end{enumerate}
Among these three experiments, D1 is the experiment where the input observation and the output to be predicted are two different physical quantities. To the best of our knowledge, none of the existing graph-signal processing approaches address such a dataset and we therefore, make comparisons only with conventional linear/kernel regression. The other two experiments D2 and D3 are performed for two reasons. The first reason is to compare our method against the kernel-ridge regression (KRR) method of \cite{kergraph1,kergraph4}. In these experiments, the input and the output both lie on the same graph and are the same physical quantities, making it applicable to the KRR method. We choose KRR as the competing method in these two experiments because it has been claimed to provide a state-of-the-art performance \cite{kergraph1}. The second reason is to investigate the performance of our method when we simultaneously learn an underlying graph from the available data.  
%The reason to perform four experiments is to show competitive performance of proposed method in diverse applications.
	
	We use the normalized-mean-square-error (NMSE) as the measure of the prediction performance:
\begin{equation}
	\mbox{NMSE}=10\log_{10}\left(\frac{\mathrm{E}\|\mathbf{Y}-\mathbf{T}_0\|_F^2}{\mathrm{E}\|\mathbf{T}_0\|_F^2}\right),\nonumber
	\end{equation}
	where $\mathbf{Y}$ denotes the regression output matrix and $\mathbf{T}_0$ the true value of target matrix, meaning that $\mathbf{T}_0$ does not contain any noise. The expectation operation $\mathrm{E}(\cdot)$ is realized as the sample average over multiple experiment trials.
	In the case of real-world examples, we compare the performance of the following four regression approaches:
	\begin{enumerate}
		\item linear regression (LR): ${k}_{m,n}=\pmb\phi(\mathbf{x}_m)^\top\pmb\phi(\mathbf{x}_n)$, where $\pmb\phi(\mathbf{x})=\mathbf{x}$ and $\beta=0$,
		\item linear regression over graphs (LRG): ${k}_{m,n}~=~\pmb\phi(\mathbf{x}_m)^\top\pmb\phi(\mathbf{x}_n)$ and $\beta\neq0$, where $\pmb\phi(\mathbf{x})=\mathbf{x}$,
		\item kernel regression (KR): Using $\beta=0$ and the radial basis function (RBF) kernel ${k}_{m,n}=\displaystyle\exp\left(-\frac{\|\mathbf{x}_m-\mathbf{x}_n\|_2^2}{\sigma^2\sum_{m',n'}\|\mathbf{x}_{m'}-\mathbf{x}_{n'}\|_2^2/N}\right)$, and
		\item kernel regression over graphs (KRG): Using $\beta\neq0$ and the RBF kernel ${k}_{m,n}=\displaystyle\exp\left(-\frac{\|\mathbf{x}_m-\mathbf{x}_n\|_2^2}{\sigma^2\sum_{m',n'}\|\mathbf{x}_{m'}-\mathbf{x}_{n'}\|_2^2/N}\right)$.
	\end{enumerate}
	The regularization parameters $\alpha$, $\beta$, $\sigma^2$ and $\nu$ for different training data sizes are found through a {\color{black} five-fold crossvalidation on the training set. While performing the crossvalidation, we assume that clean target vectors are available.}  We wish to emphasize that our goal here is to illustrate that LRG and KRG are better than LR and KR, respectively. Depending on the choice of the kernel used, one kernel may perform better than the other and that there is generally no guarantee that a Gaussian kernel always outperforms the linear kernel in practice. {\color{black}As regards determining which kernel is best suited to an application, there is often no direct answer than a trial-and-error approach of using different kernels and comparing their performance. An alternative is to use a kernel selection approach in the form of multi-kernel regression \cite{multikernel_5,multikernel_Arun,kergraph2} which `selects' the best kernels from a bag of kernels.}
		
\subsection{Regression for synthesized data}
{\color{black}We perform regression for the synthesized dataset where the target vectors to be predicted are smooth over a specified graph. 
	%Our hypothesis is that KRG will outperform KR when the training data is limited in size and corrupted with noise. To verify the hypothesis, 
	We generate synthesized data where we know the ground truth. 
	A part of the data is used for the training in presence of additive noise, and our task is to predict for the remaining part of the data, given the information about correlations in form of kernels.} 
In order to generate the synthesized data, we use random small-world graphs from the Erd\H{o}s-R\'{e}nyi and the Barab\'{a}si-Albert models \cite{Newman} with the number of nodes equal to $M=50$. {\color{black}We generate a total of $S$ target vector realizations. We adopt the following data generation strategy: We first pick $M$ independent vector realizations from an $S$-dimensional Gaussian vector source $\mathcal{N}(\mathbf{0}, \mathbf{C}_{S})$, where $\mathbf{C}_{S}$ is an $S$-dimensional covariance matrix drawn from the inverse Wishart distribution with an identity hyperparameter matrix. We use a highly correlated covariance matrix $\mathbf{C}_{S}$ such that each $S$-dimensional vector has strongly correlated components. Thus, we create a data matrix with $M$ columns such that each column is an $S$-dimensional Gaussian vector. Each row vector of the data matrix has a size $M$ and the row vectors of the data matrix are correlated to each other. We denote a row vector by $\mathbf{r}$. We then select the row vectors $\{ \mathbf{r} \}$ one-by-one and project them onto the specified graph to generate target vectors $\{\mathbf{t}\}$ that are smooth over the graph while maintaining the correlation between observations, by solving the following optimization problem:
	\begin{eqnarray}
	\mathbf{t}=\arg\min_{\mathbf{z}} \left\{ \|\mathbf{r}-\mathbf{z}\|_2^2+ \mathbf{z}^\top\mathbf{Lz} \right\}. \nonumber
	\end{eqnarray}
	We randomly divide the $S$ data samples into training and test sets of equal size $N_{tr}=N_{ts}=S/2$. We define the kernel function between the $i$th and $j$th data samples $\mathbf{z}_i$ and $\mathbf{z}_j$ to be
		\begin{equation*} {k}_{i,j}\triangleq\mathbf{C}_S(i,j),
		\end{equation*} considering the same kernel for all the graph nodes. The choice of the kernel is motivated by the assumed generating model. Given the training set of size $N_{tr}$, we choose a subset of $N$ data samples to make predictions for the $N_{ts}$ test data samples using the kernel regression over graphs. }
	The training target vectors are corrupted with additive white Gaussian noise at varying levels of signal-to-noise ratio (SNR). We repeat our experiments over 100 realizations of the graphs and noise realizations. We compare the performance of KRG with KR. We observe from Figure \ref{ERgraph1} that for a fixed training data size of $N=50$, KRG outperforms KR by a significant margin at low SNR levels (below 10dB). As the SNR-level increases, the NMSE of KRG and KR almost coincide. A similar trend is also observed in the case of Barab\'{a}si-Albert graphs, which is not reported for brevity. In Figure \ref{BAgraph}, we show the NMSE obtained with KRG and KR on both the graph models, as a function of the training data size at an SNR-level of $5$ dB. We observe that KRG consistently outperforms KR and that the gap between the NMSE of KRG and KR reduces as the training data size increases. The results shown in Figure \ref{ERgraph1} and \ref{BAgraph} verify our hypothesis.
	%
	%and \ref{BAgraph} that KRG outperforms KR by a significant margin.
	%The trend is similar for large dimensional graphs as well and is not reported here to avoid repetition.  

	\begin{figure}
		\centering
		$
		\begin{array}{cccc}
		\subfigure[\hspace{-.2in}]{\includegraphics[width=1.7in]{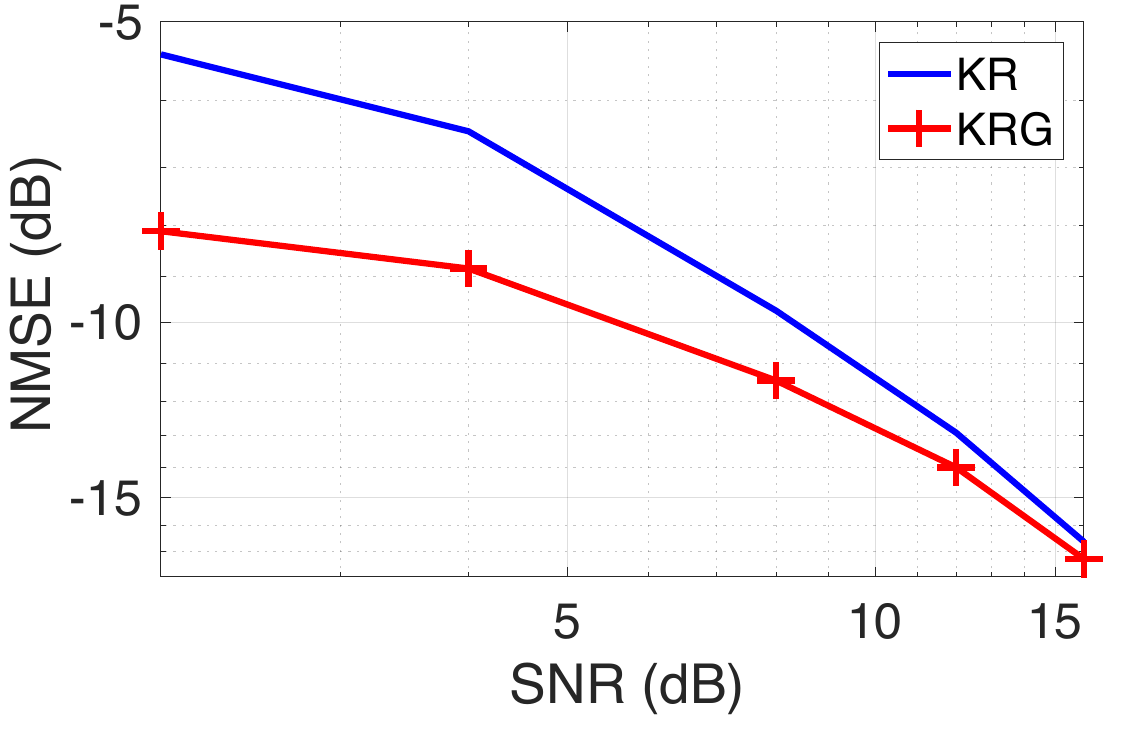}}\,\,
		\subfigure[\hspace{-.2in}]{\includegraphics[width=1.7in]{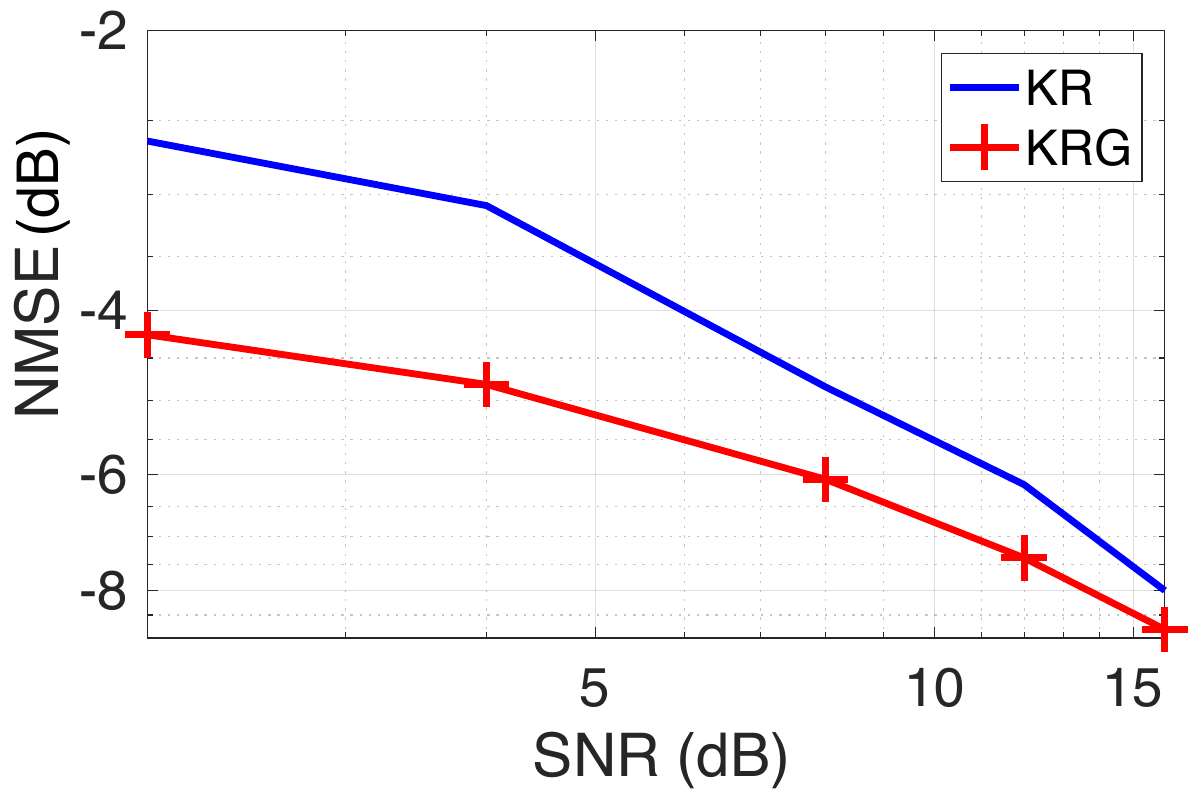}}
		\end{array}
		$
		\caption{ Performance for the synthesized dataset using Erd\H{o}s-R\'{e}nyi graphs. We plot the NMSE against SNR for the training data size $N=50$. (a) NMSE for training data, and (b) NMSE for test data.
			%: (a) for training data, and (b) for testing data.  Normalized MSE at 5 dB SNR level: (c) for training data, and (d) for testing data.
		}
		\label{ERgraph1}
	\end{figure}
	\begin{figure}
		\centering
		$
		\begin{array}{ccc}
		\subfigure[\hspace{-.1in}]{\includegraphics[width=1.7in]{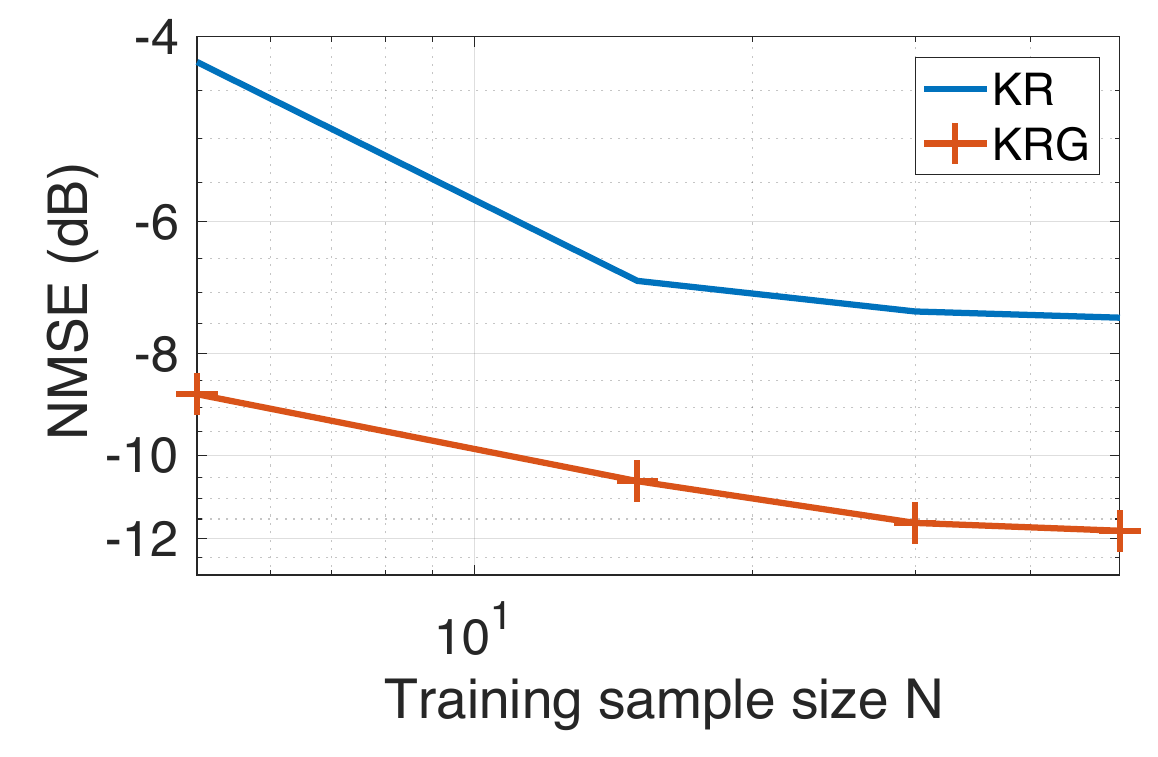}}\hspace{.in}
		\subfigure[\hspace{-.1in}]{\includegraphics[width=1.7in]{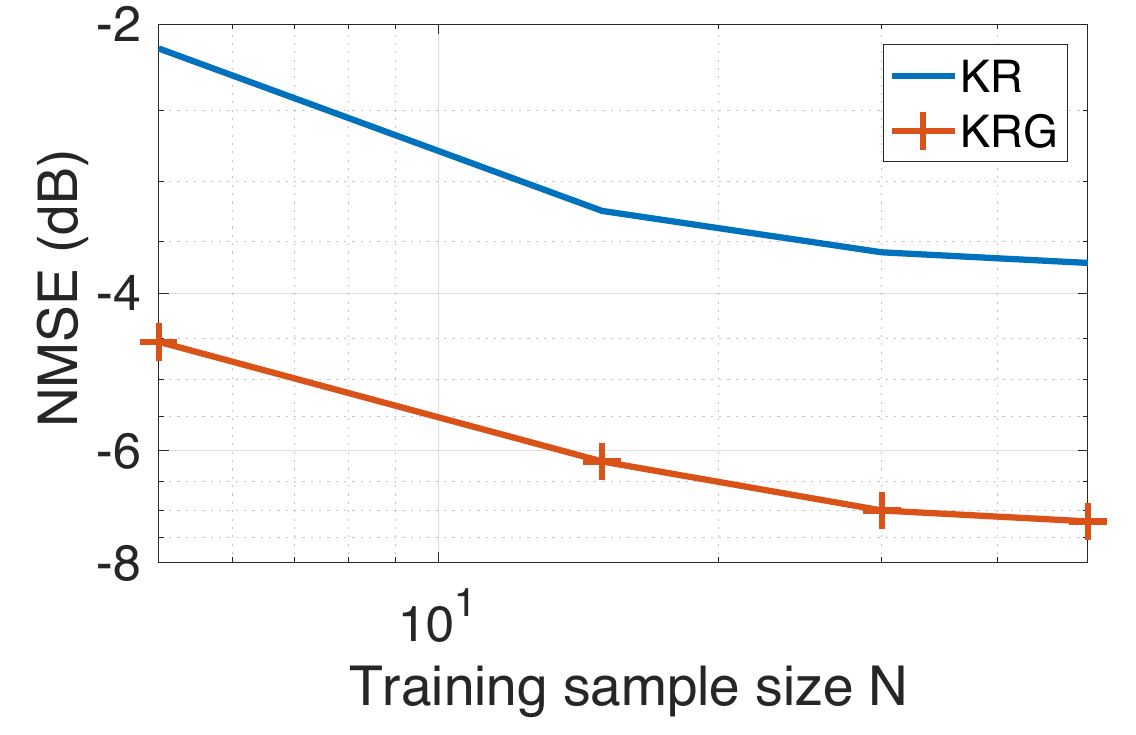}}\hspace{-.0in}\\
		\subfigure[\hspace{-.1in}]{\includegraphics[width=1.7in]{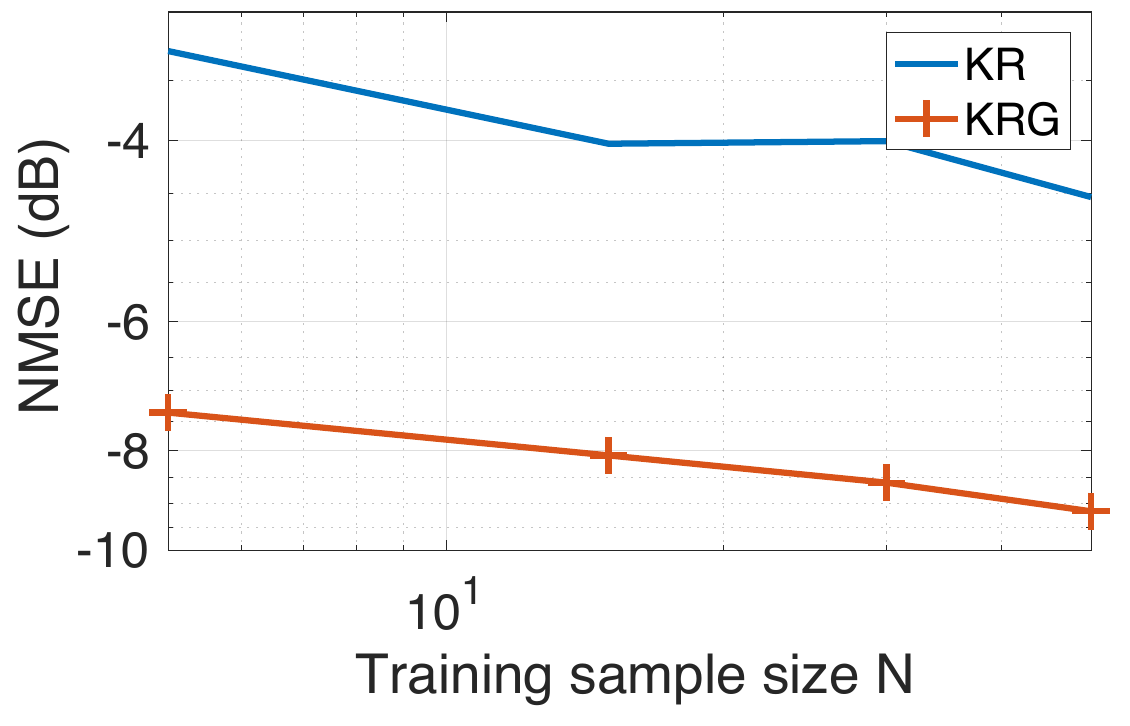}}\hspace{.in}
		\subfigure[\hspace{-.1in}]{\includegraphics[width=1.7in]{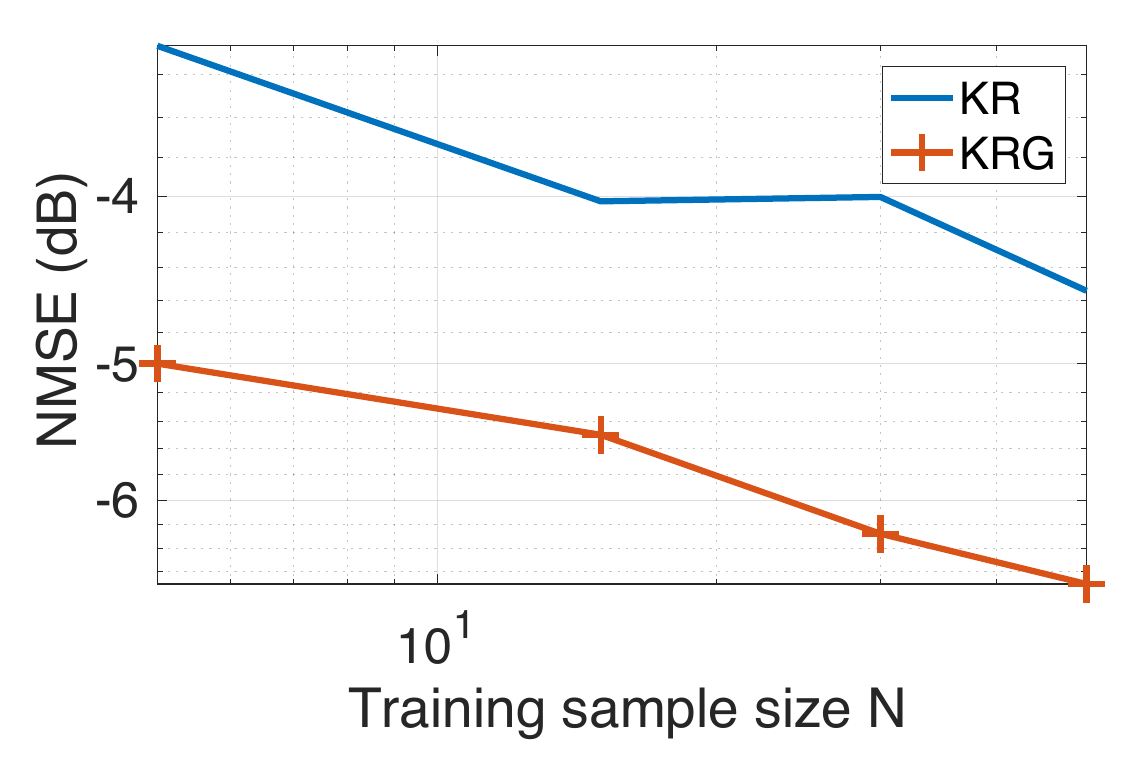}}\hspace{-.0in}
		\end{array}
		$
		\caption{Performance for the synthesized dataset at a 5 dB SNR level. We use Erd\H{o}s-R\'{e}nyi graphs for subfigures (a) and (b). (a) NMSE for training data, and (b) NMSE for test data. Then we use Barab\'{a}si-Albert graphs for subfigures (c) and (d). (c) Training data performance and (d) Test data performance.  
			%(c) Normalized MSE for training data, and (d) testing data at $0$dB SNR level.
		}
		\label{BAgraph}
	\end{figure}

	{\color{black}
		\subsection{Experiment D1: Prediction of the temperature of the cities using air-pressure observations}	
			\begin{figure}[t]
			\centering$\begin{array}{cc}
			\subfigure[]{\includegraphics[height=1.8in]{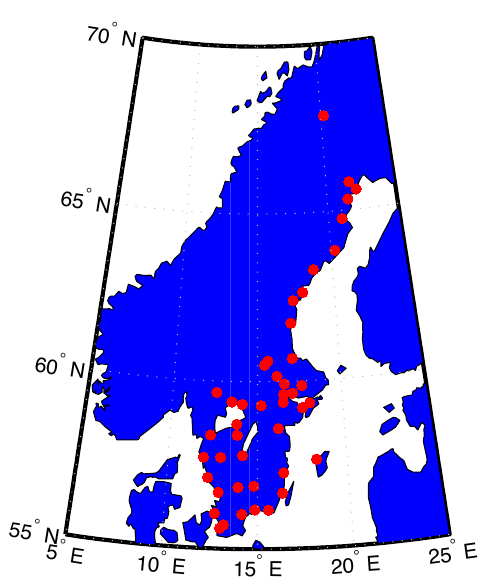}}\hspace{.in}
			\subfigure[]{\includegraphics[height=1.8in]{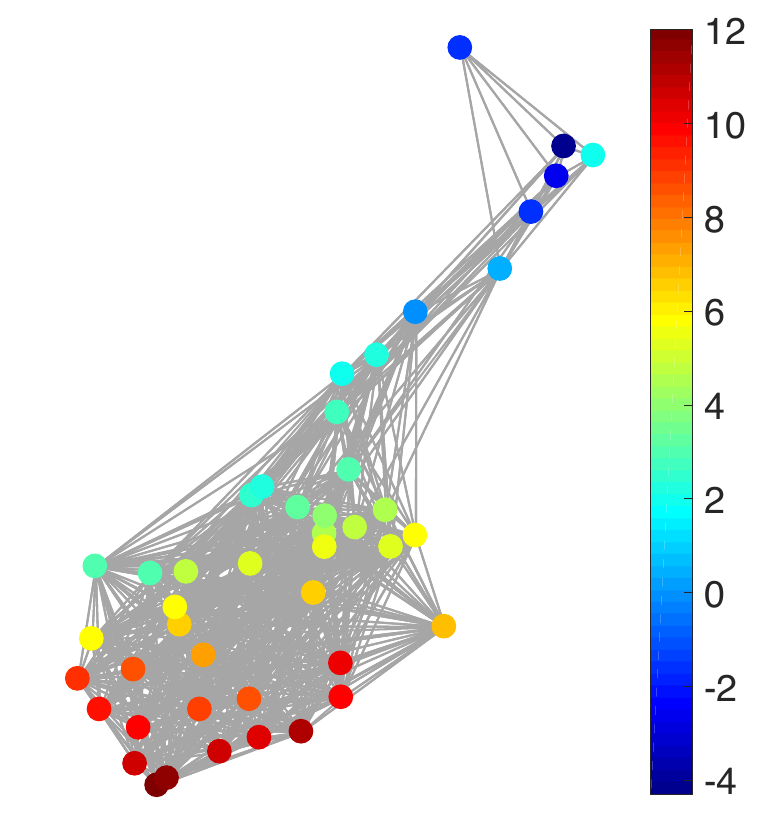}}\hspace{-.0in}
			\\
			\subfigure[]{
				\includegraphics[width=2.2in]{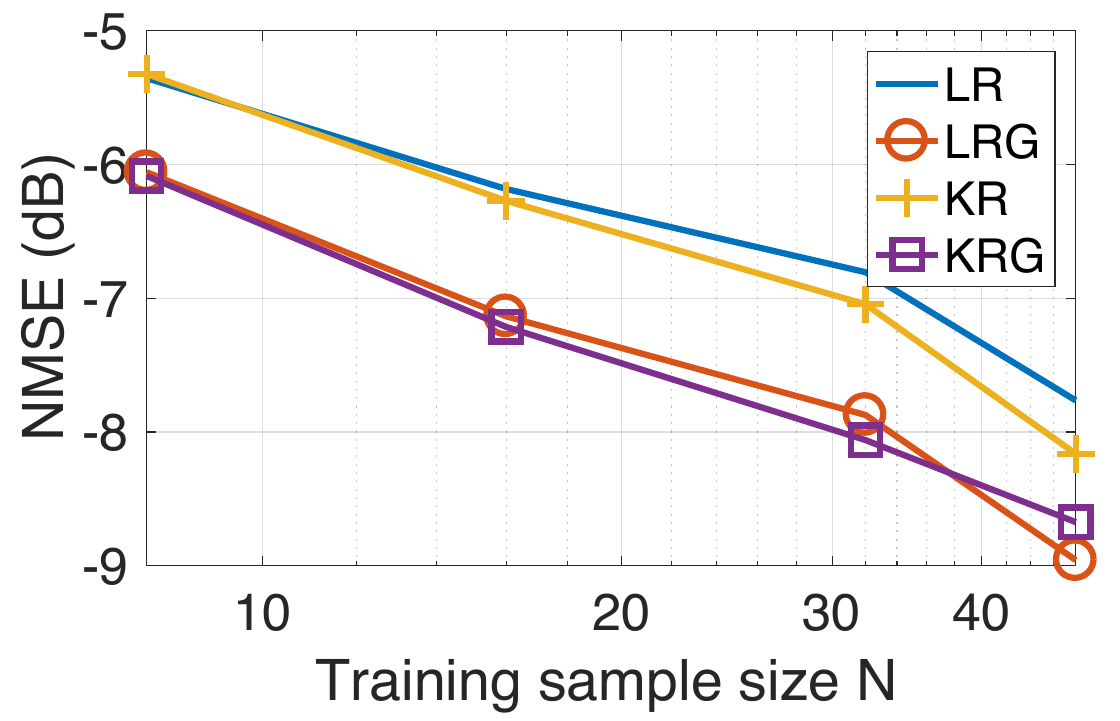}}\\
			%\subfigure[]{
				%\includegraphics[width=2.2in]{krg_smhi18_nmse_large.pdf}}
			\end{array}
			$
			\caption{Results for the experiment (D1). (a) The map of Sweden with the major 45 cities indicated, (b) {\color{black}True temperature measurement signal for a particular day in the dataset shown in units of degree celsius (The graph is the same as that in subfigure (a) but without the underlying map.)}. (c) NMSE as a function of the training sample size, with additive white Gaussian noise at a $10$ dB SNR-level.
				% and (d) with larger perturbations at two of the nodes.
			}
			\label{fig:smhi18}
		\end{figure}
		
We now consider the experiment where the input and the output of the kernel regression are two different physical quantities. The task is to predict the temperature of the cities in Sweden from the air pressure observations at those cities. We predict the temperature as the average daily temperature for 24 hours of a day; the input consists of the air pressure observations collected on an hourly basis for the same day. 
%Our hypothesis remains the same as with the previous experiment  --  that KRG would outperform KR when the training data is limited in size and corrupted with noise.		
		
		For the experiment, we collected the temperature and air-pressure measurements from the 25 most populated cities in Sweden. The data was collected for a period of two months from February to March of 2018.  In Figure \ref{fig:smhi18}(a), we indicate the 45 most populated cities in Sweden. We consider 25 of these 45 cities in this experiment since the relevant data was not available at the remaining cities. The data is available publicly from the Swedish Meteorological and Hydrological Institute \cite{SMHI}. We predict the temperature of the 25 cities as the output vector or the target. The input is taken to be the air pressure measurement at all those 25 cities collected on hourly basis. This results in an input vector with $24\times 25=600$ components, which means that we have $\mathbf{t}_n\in\mathbb{R}^{25}$ and $\mathbf{x}_n\in\mathbb{R}^{600}$. The data from the first 48 days is taken as the training set, and the data from the remaining 12 days is used for testing.
		Let $d_{ij}$ denote the geodesic distance between the $i$th and $j$th cities {\color{black}in kilometres}, $\forall i,j\in\{1,\cdots, 25\}$. Then, we construct the adjacency matrix $\mathbf{A}$ for the graph by setting 
		\begin{equation}
		\mathbf{A}(i,j)=\exp{\left(-\frac{d_{ij}^2}{\sum_{i,j}d_{ij}^2}\right)}.\nonumber
		\end{equation}
		%For each $N$, we compute the NMSE by averaging over 50 different random training subsets of $N$ drawn from full training set of size $N_{tr}=48$. 
		
In our experiments, we randomly sample for the training observations from the full training set for various $N \leq N_{tr}=48$.  We consider the case when the training targets are corrupted with additive white Gaussian noise at a 10 dB SNR level. We compare the performances of LR, LRG, KR, and KRG. For the test data, the NMSE as a function of the training sample size $N$ is shown in Figure \ref{fig:smhi18}. We observe that LRG and KRG outperform LR and KR, respectively, by a significant margin, particularly at smaller training sample sizes. In addition to this, we find that KRG outperforms LRG for this experiment, though this is not necessarily guaranteed for all the datasets and under all experimental conditions. 
	\begin{figure}[t]
		\centering
		$
		\begin{array}{cc}
		\hspace{-.1in}
		\subfigure[\hspace{-.2in}]{\includegraphics[width=2.4in]{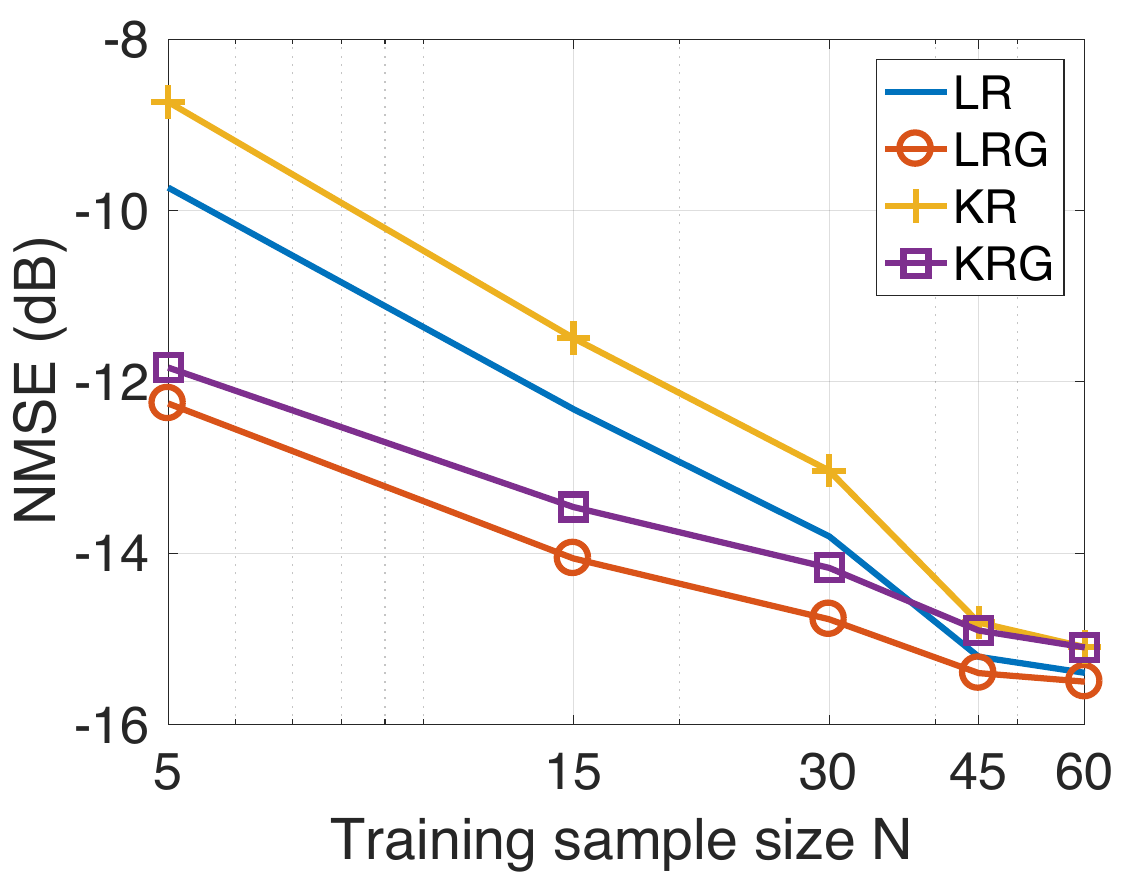}}\hspace{-.0in}\\
			\hspace{-.05in}
		\subfigure[\hspace{-.1in}]{\includegraphics[width=2.4in]{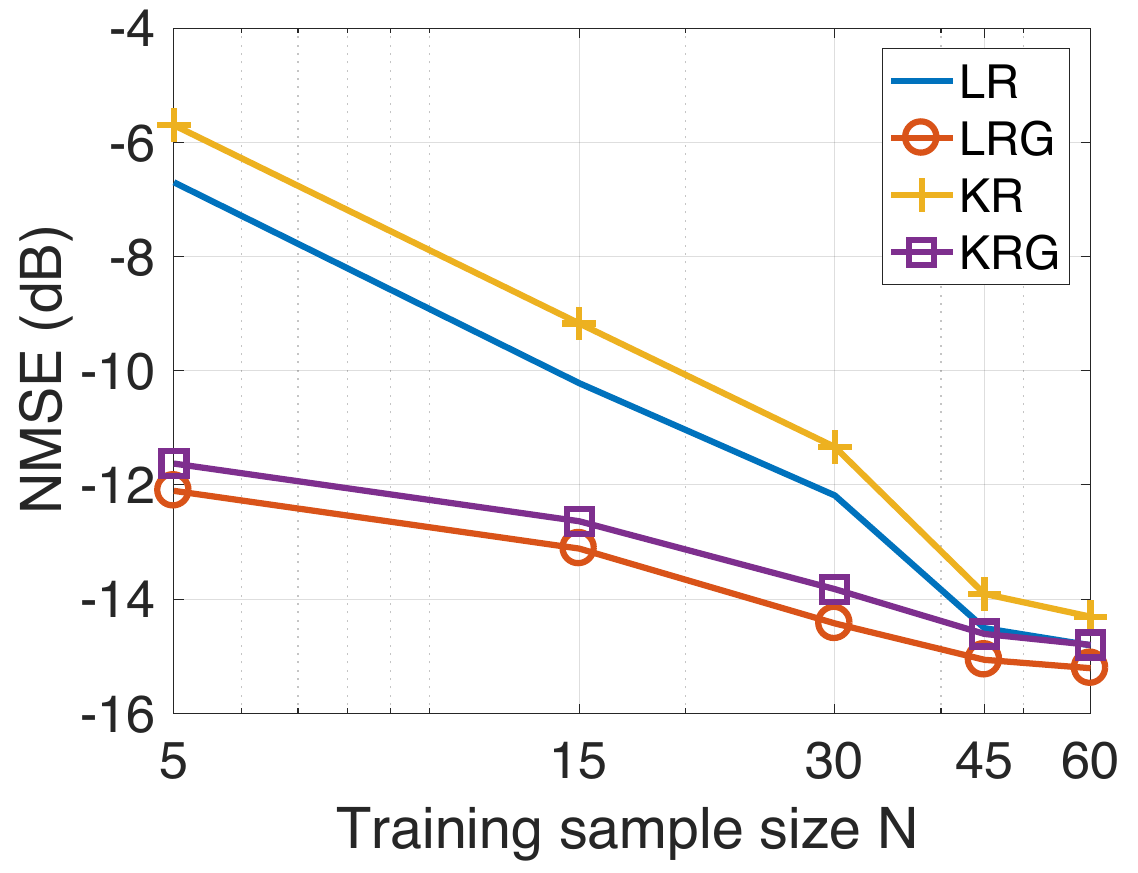}}\hspace{-.0in}
	%	\subfigure[\hspace{-.1in}]{\includegraphics[width=2.3in]{fig3e.pdf}}\hspace{-.0in}
	%	\subfigure[\hspace{-.1in}]{\includegraphics[width=2.3in]{fig3f.pdf}}\hspace{-.0in}
		\end{array}
		$
		\caption{ Results for experiment (D2).  (a) NMSE for test data with additive white Gaussian noise at a $5$dB SNR level, (b) NMSE for test data with additive white Gaussian noise at a $0$dB SNR level.
			% (c) NMSE  test data at with missing samples, and (d) NMSE  test data at with large perturbations.
		}
		\label{swedetempgraph}
	\end{figure}
	}

	\subsection{Experiment D2: Temperature prediction from the current day to the next day}
	\label{subsec:Temperature_prediction_for_cities_in_Sweden}

	\begin{figure}[t]
		\centering
		$
		\begin{array}{cccc}
		%\subfigure[]
		{\includegraphics[width=2.4in]{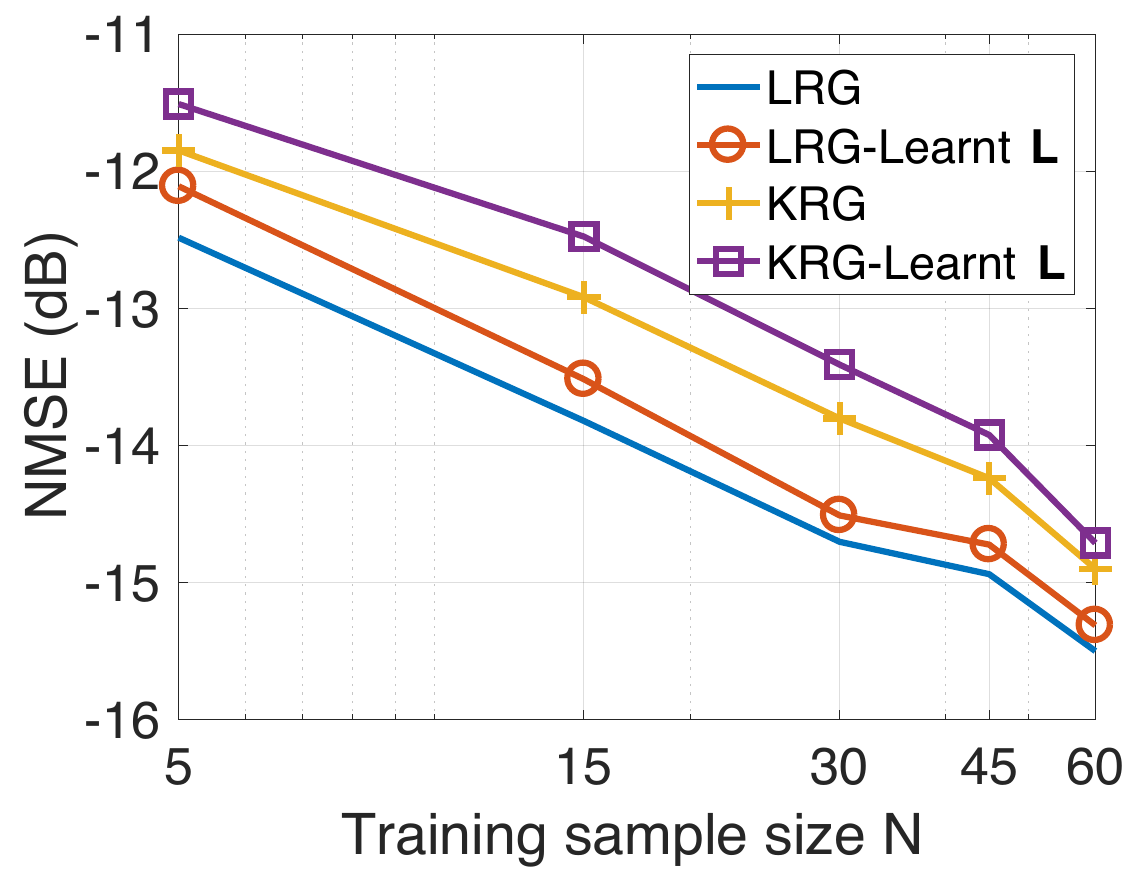}}\hspace{-.0in}
		%	\subfigure[]{\includegraphics[width=1.7in]{mse_temp17_learn_nonzeroL_5db.pdf}}\hspace{-.0in}
		%\subfigure[]{\includegraphics[width=1.6in]{fig4b.pdf}}\hspace{.in}
		%\subfigure[]{\includegraphics[width=1.6in]{fig4c.pdf}}\hspace{-.0in}
		%\subfigure[\hspace{-.1in}]{\includegraphics[width=1.6in]{fig4d.pdf}}\hspace{-.0in}
	%	\subfigure[\hspace{-.1in}]{\includegraphics[width=2in]{fig4e.pdf}}\hspace{-.0in}
		\end{array}
		$
		\caption{\color{black} Results for experiment (D2) with graph learning, for training data samples at a $5$dB SNR-level for $N=45$.
			%, and (b) when initialized with $\mathbf{L}$ corresponding to the geodesic graph. %Geodesic Laplacian, (c) Estimated Laplacian from LRG, (d) Estimated Laplacian from KRG.
			% and (e) vectorized entries of the different Laplacians. The Laplacians have been scaled to have maximum eigenvalue equal to unity.
		}
		\label{swedetempgraphlearn2}
	\end{figure}
\begin{table*}[!h]
	\centering
	\caption{{\color{black}Performance comparison of the proposed methods and KRR for the experiment D2 (using NMSE in $\mathrm{d}$B)}}
	\begin{tabular}{|c|c|c|c|c|c|c|c|}
		\hline
		
		\multicolumn{1}{|c|}{} &
		\multicolumn{3}{c|} {Kernel Ridge Regression (KRR)} &
		\multicolumn{4}{c|}{Proposed Methods} \\ \cline{2-8}	
		
		Training sample& Diffusion & Covariance kernel & Covariance kernel&LRG at &KRG at &LRG at &KRG at \\
		size $N$& kernel&at 5dB SNR& at 0dB SNR&5dB SNR&5dB SNR&0dB SNR&0dB SNR\\
		\hline
		5& 2.8$\times 10^{-4}$& -6.5& -3.5&-12.3&-11.9&-12.1&-11.6\\
		15&''&-7.2&-4.9&-14.0&-13.4&-13.1&-12.7\\
		30&''&-9.8&-6.7&-14.8&-14.2&-14.4&-13.8\\
		45 &''&-10.5&-7.8&-15.4&-14.9&-15.0&-14.6\\
		60&''&-12.2&-9.5&-15.5&-15.1&-15.2&-14.8\\
		\hline
	\end{tabular}
	\label{tab:krr_temp}\end{table*}

	%	\begin{figure}[t]
	%		\centering
	%		$
	%		\begin{array}{ccc}
	%		\hspace{-.0in}
	%	%	\subfigure[]{\includegraphics[width=2.0in]{./Codes/Figures/SwedeTemp/trainlearn0dB.eps}}\hspace{-.in}\\
	%		\subfigure[]{\includegraphics[width=1.8in]{./Codes/Figures/SwedeTemp/testlearn10dB.eps}}\hspace{-.0in}
	%		\end{array}
	%		$
	%		\caption{ For the temperature graph of 45 cities in Sweden with graph learning (a) Normalized MSE for testing data at $5$dB SNR level.
	%		}
	%		\label{swedetempgraphlearn}
	%	\end{figure}
	%	%
	%
	%
	
{\color{black} 	In this experiment, the task is to predict the temperature of several Swedish cities for the next day from the temperature observations of the current day. For the experiment, we consider the temperature measurements from the 45 most populated cities in Sweden taken for a period of three months from September to November 2017.
	 %The hypothesis is once again that our approach brings in a significant performance gain in noisy and limited training data scenarios. 
	 Since both the input and the output are temperatures, this experiment represents a graph signal reconstruction/recovery problem and hence, we compare our method with the KRR method. We have already mentioned that KRR is a state-of-the-art method in graph signal recovery\cite{kergraph1,kergraph4}. Further, we also consider the case when the underlying graph is not known a-priori. In this case, we learn an underlying graph and compare the performance of our approach against the case where the graph is known a-priori.} The data is available publicly from the Swedish Meteorological and Hydrological Institute \cite{SMHI}. The cities are indicated in the map of Sweden in Figure \ref{fig:smhi18}(a). We consider the target vector $\mathbf{t}_n$ to be the temperature measurement of a particular day, and $\mathbf{x}_n$ to be the temperature measurements {\color{black} (in degree celsius units)} from the previous day. {\color{black} We have 90 input-target data pairs in total, divided into the training set and the test set of sizes $N_{tr}=60$ and $N_{ts}=30$, respectively.}
	%a part of the data is used for training and the rest for testing. 
	%We report the results by averaging over 100 experiments where each experiment uses a random partition of the data into training and testing datasets. 
	%Let $d_{ij}$ denote the geodesic distance between cities $i$ and $j$ {\color{black}in kilometres}, $\forall i,j\in\{1,\cdots, 45\}$. We construct the adjacency matrix by setting 
	%\begin{equation}
	%\mathbf{A}(i,j)=\exp{\left(-\frac{d_{ij}^2}{\sum_{i,j}d_{ij}^2}\right)}.\nonumber
	%\end{equation}
	%In order to remove self loops, the diagonal of $\mathbf{A}$ is set to zero.  We simulate noisy training data by adding zero-mean white Gaussian noise to the targets. %The Laplacian distribution is heavy-tailed and simulates impulsive noise at various weather stations. 
	Once again, we consider the geodesic distance based graph.
	For each training dataset size $N$, we compute the NMSE by averaging over 50 different random training subsets of size $N$ drawn from the full training set of size $N_{tr}$. 
	In Figure \ref{swedetempgraph}, we show the NMSE for the test set at SNR-levels of $5$ dB and $0$ dB. We observe that KRG outperforms other regression methods by a significant margin, particularly at low sample sizes $N$. Next, we compare our methods with KRR.  
	
%	{In order to simulate a more realistic noisy training scenario, we further consider two different cases of noisy training: {\it missing data} and {\it large perturbations}. In the case of missing data experiment, we zero out the target values at randomly chosen five nodes for each training sample and apply the various regression approches. In the perturbation experiment, we chose five of the nodes randomly for each training sample and perturb the signal value to five times of the original values. The NMSE obtained for LR, LRG, KR, and KRG as a function of $N$ is shown in Figures \ref{swedetempgraph}(a)-(d). Thus, we observe that graph-structure improves robustness of regression in terms of prediction performance for LR and KR.} 
	{\color{black}
\subsubsection{Comparison with KRR}	

KRR deals with a sub-sampling problem where the signal values are predicted at a set of nodes from the signal values given at the remaining set of nodes. Therefore, we formulate the one-day temperature prediction problem in a suitable sub-sampling setup where KRR can be used. In the sub-sampling setup, KRR minimizes the following convex cost:
\begin{equation}
\arg\min_{\pmb\alpha} \left\|\mathbf{x}-\mathbf{S}\bar{\mathbf{K}}\mathbf{S}\pmb\alpha\right\|_2^2+\mu\pmb\alpha^\top\bar{\mathbf{K}}\pmb\alpha,\mbox{  s.t.  } \bar{\mathbf{K}}\pmb\alpha=\left[\begin{matrix}
\mathbf{y}\\
\hat{\mathbf{x}}
\end{matrix}\right]
\end{equation}		
where $\mathbf{x} \in \mathbb{R}^s$ is the input observation signal corresponding to a subset $\Omega$ with $s$ nodes, from the total set of $M$ nodes. Here, $\mathbf{S}$ denotes the sampling matrix obtained by concatenating the zero matrix and the $s$-dimensional identity matrix; $\bar{\mathbf{K}}$ is the kernel matrix across all nodes of the graph, and $\pmb\alpha$ is the vector of KRR coefficients, and $\hat{\mathbf{x}}$ is the estimate of the graph signal produced by KRR at $\Omega$. The estimate of the entire graph signal is then given by:
\begin{equation}
\left[\begin{matrix}
\mathbf{y}\\
\hat{\mathbf{x}}
\end{matrix}\right]=\bar{\mathbf{K}}\mathbf{S}^\top(\mathbf{S}\bar{\mathbf{K}}\mathbf{S}^\top+\mu s\mathbf{I}_s)^{-1}\mathbf{x}.
\end{equation}
Thus, KRR achieves an extrapolation of the graph signal from the nodes in $\Omega$ to those outside it using the graph topoplogy employed in the extrapolation kernel. The parameters related to the above prediction and kernels are found by cross-validation. 
In all the experiments employing KRR\cite{kergraph1,kergraph3}, we have used the same diffusion kernels and the covariance kernels used by the authors in the corresponding articles\cite{kergraph1,kergraph3}. 
%For covariance kernels, we use the sample covariance as the estimate of the real covariance matrix as also employed in \cite{kergraph1}. 
%We also note that whereas the performance of the covariance kernel depends on the number of training samples used to obtain the covariance estimate, the diffusion kernel based KRR does not use training samples. 
%As with  KRG, we obtain the hyperparameter $\mu$ and $\sigma^2$ for KRR by cross-validation. 

Since we consider the one-day temperature prediction problem, we use the space-time variant of the KRR proposed in \cite{kergraph3}, by taking the adjacency matrix given by the Cartesian product of the geodesic graph $\mathbf{A}$ and the temporal dynamics graph for one time step $\mathbf{B}=\left[\begin{matrix}
	0 &1\\1&0
	\end{matrix}\right]$, meaning that each node at time $n$ is connected to the corresponding node at time $n+1$ by an edge with unity weight. The composite or augmented graph\cite{kergraph3} for the two days is then given by the Cartesian product $\mathbf{A}\oplus\mathbf{B}=
	\left[\begin{matrix}
	\mathbf{A}&\mathbf{I}\\
	\mathbf{I}&\mathbf{A}
	\end{matrix}\right]$.
	We observe from Table \ref{tab:krr_temp}, that our approach significantly outperforms KRR for both the covariance and the diffusion kernels. We note that the performance of the covariance kernel is better than that of the diffusion kernel, and this trend is in agreement with the results reported in \cite{kergraph3}. We also observe that the performance of KRR and our approaches improve as more training data becomes available.
	
The performance of our approach is significantly better than that of KRR. This can be attributed to two factors. The first factor is that KRR deals with an under-determined setup where the subsampling matrix has a special structure. The special structure is formed by concatenating the identity matrix and the zero matrix. This sampling matrix structure may not be well suited for sub-sampling. The second aspect or factor pertains to our approach: we use the advantage of explicit training and testing. This assumes the availability of a training dataset for our approach, whereas KRR does not have that as a requirement.

\subsubsection{Learning an underlying graph}
%The NMSE for larger sample sizes and higher SNRs are omitted as the performance of KRG almost coincides with that of conventional regression in these cases. 

We next consider the performance of our approach when the graph is also simultaneously learnt from the training data. This experiment serves the purpose of illustrating the effectiveness of our approach in inferring a graph suited to the prediction task even in the absence of a prior graph. 
 We use the alternating optimization strategy of Section~\ref{learngraph} initialized with $\mathbf{L}=\mathbf{0}$. We consider the training data at an SNR level of $5$ dB at different sample sizes. We find experimentally that the algorithm converges typically after five to ten iterations. In Figure \ref{swedetempgraphlearn2}, we plot the NMSE values obtained for the test data using both the fixed $\mathbf{L}$ based on the geodesic distances, and with the learnt graph. We observe that our approach learns a graph which provides agreeable performance even when initialized with the zero graph. This validates our intuition that the graph signal holds sufficient information to both infer both a meaningful underlying graph structure and perform target predictions. }

{\subsection{Experiment D3: Prediction for the fMRI voxel intensities of the cerebellum region}
	\begin{figure*}[t]
	\centering
	$
	\begin{array}{ccc}
	\hspace{-.2in}
	\subfigure[]{\includegraphics[width=2.4in]{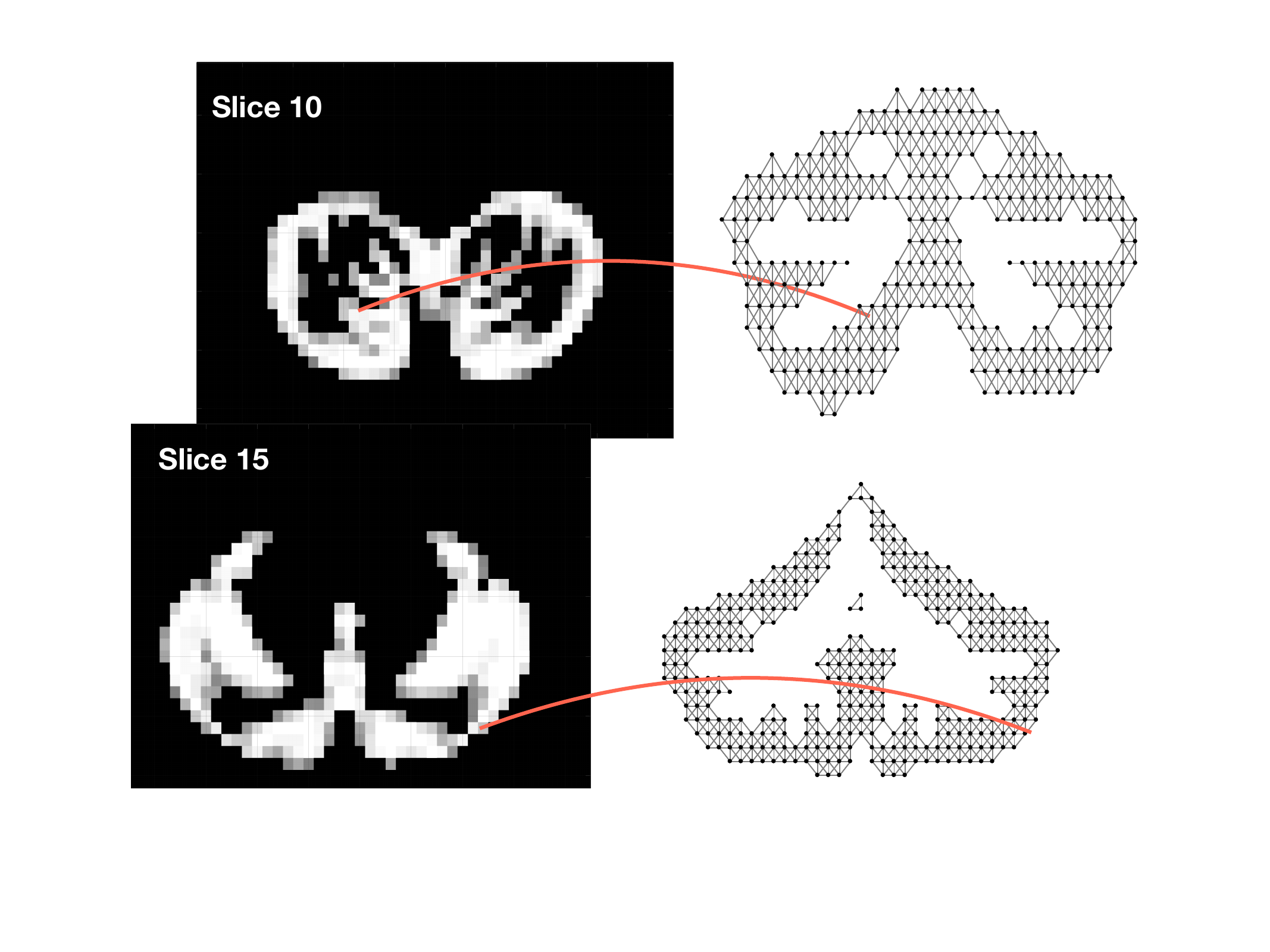}}
	\hspace{-.2in}	\subfigure[]{\includegraphics[width=2in]{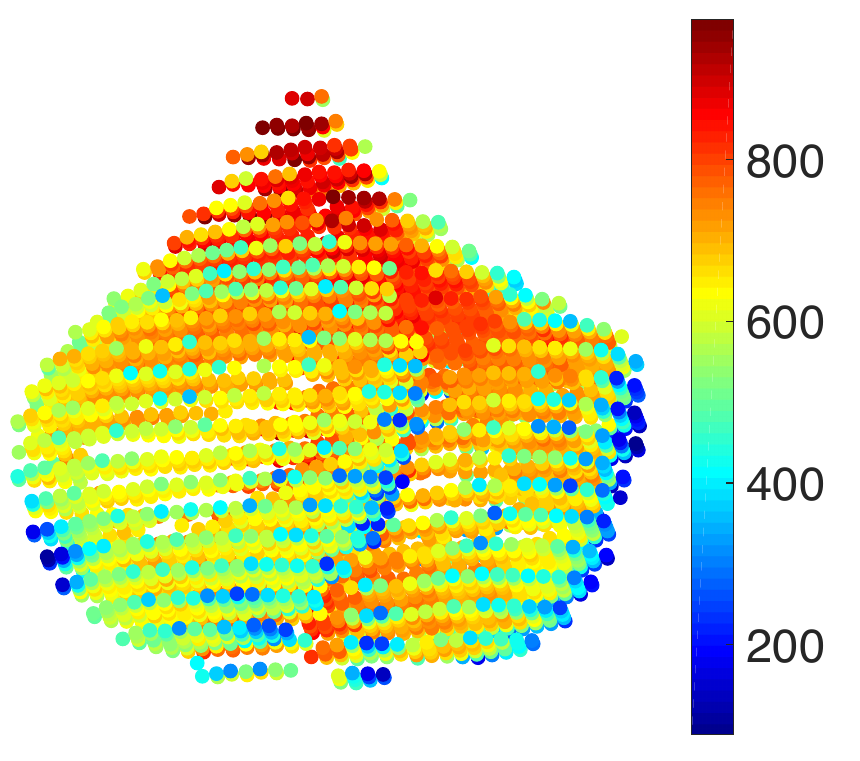}}
	\hspace{-.4in}\subfigure[]{\includegraphics[width=2in]{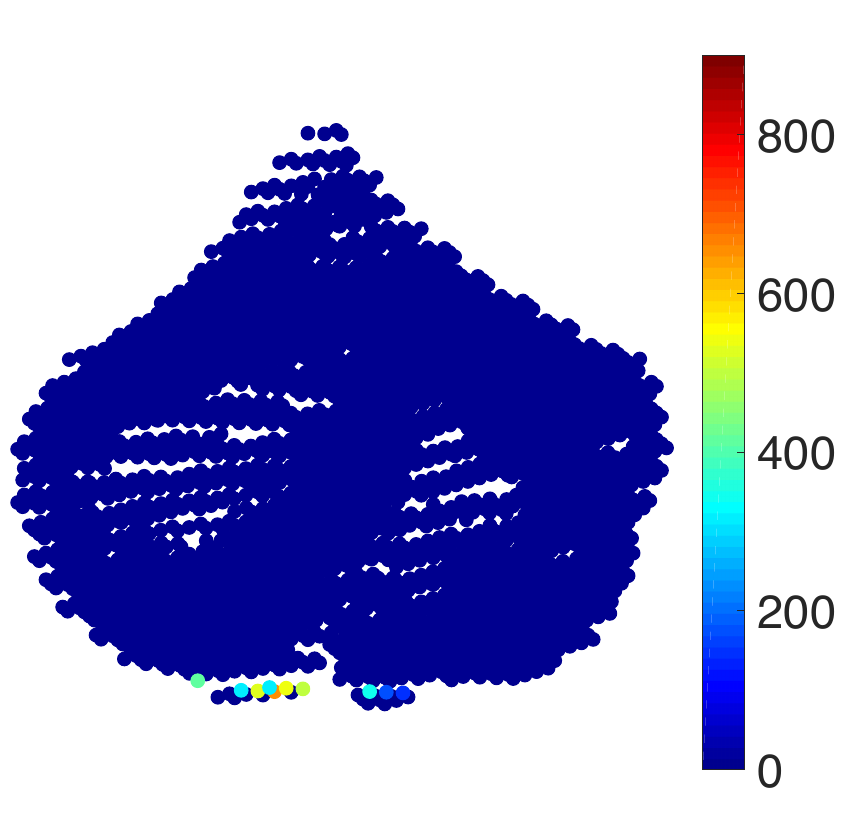}}
	\hspace{-.4in}	\subfigure[]{\includegraphics[width=2in]{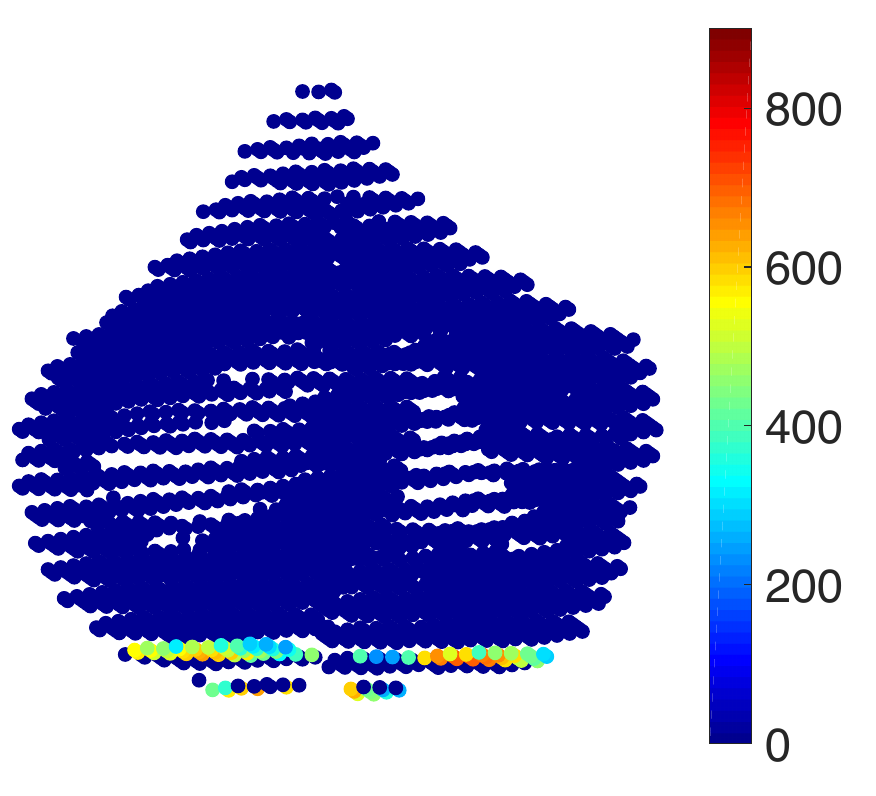}}\\
	\subfigure[]{\includegraphics[width=2in]{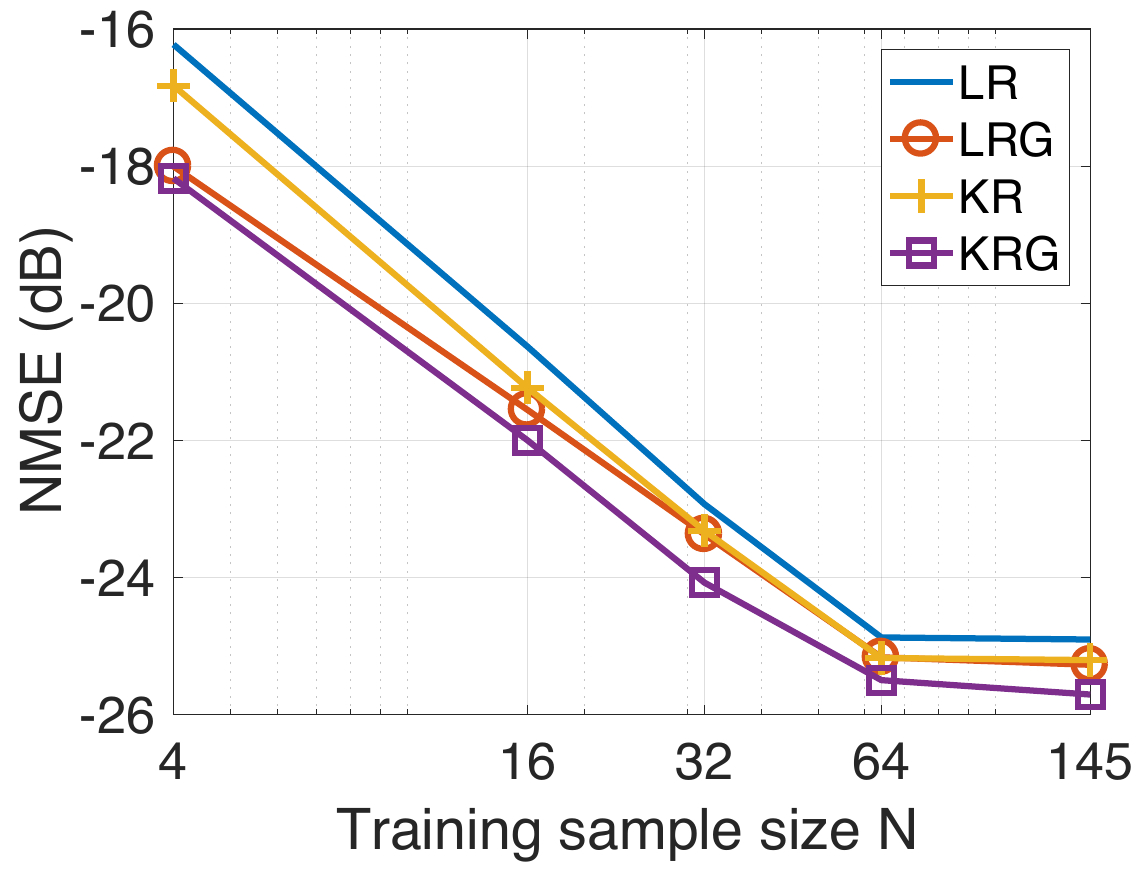}}
	\subfigure[\hspace{-.2in}]{\includegraphics[width=2in]{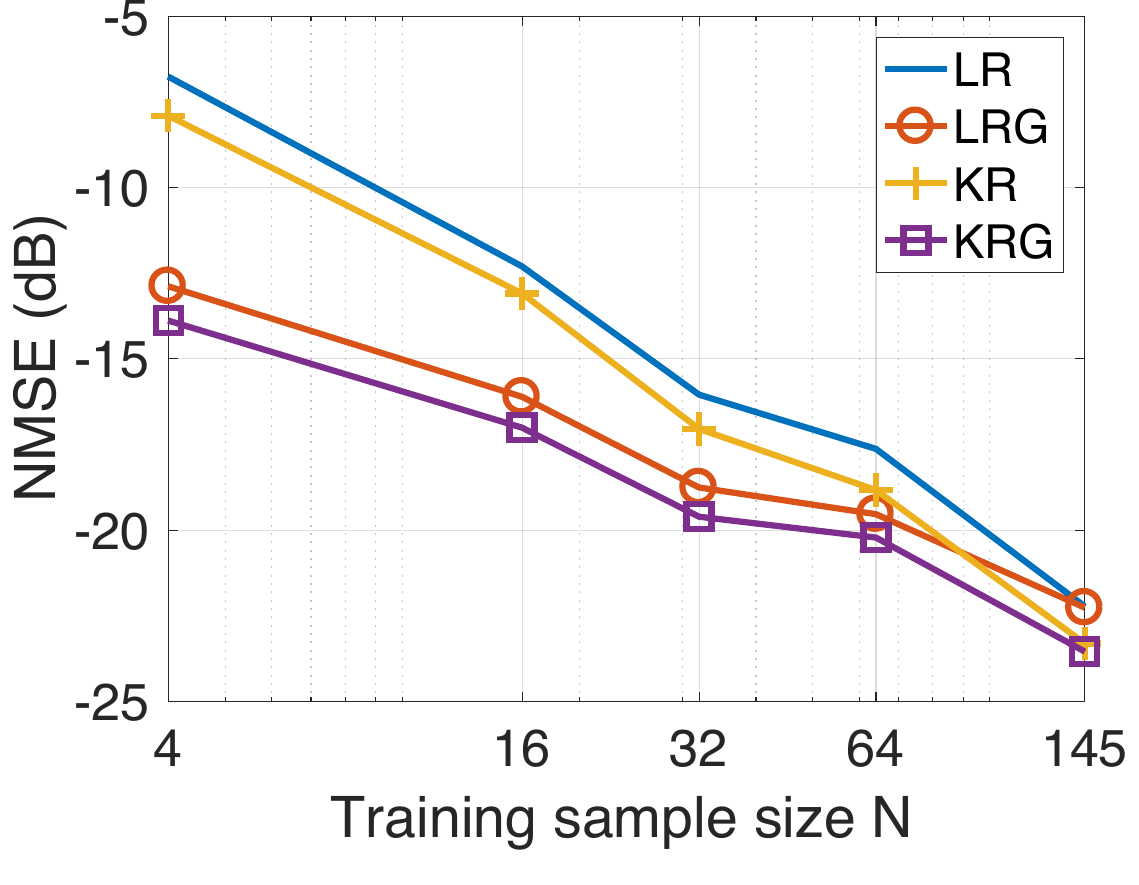}}\hspace{-.0in}
	%	\subfigure[\hspace{-.2in}]{\includegraphics[width=2in]{fig7c.pdf}}\hspace{-.0in}
	%\subfigure[\hspace{-.2in}]{\includegraphics[width=1.7in]{fig7d.pdf}}\hspace{-.0in}
	\end{array}
	$
	\caption{Results for the cerebellum data (D3). (a) Representation of how the graph is constructed from the voxels at the different slices (b) The entire graph with an instance of the graph signal, and the corresponding intensities (c) at only the voxels used as the input (d) at only the voxels used as the output (the edges are omitted for clarity),
		(e) NMSE for test data at a $10$dB SNR-level, 
		(f) NMSE for test data at a $0$dB SNR-level.
	}
	\label{cere}
\end{figure*}
	\begin{figure}[h]
	\centering
	$
	\begin{array}{cccc}
	\subfigure[]{\includegraphics[width=2.4in]{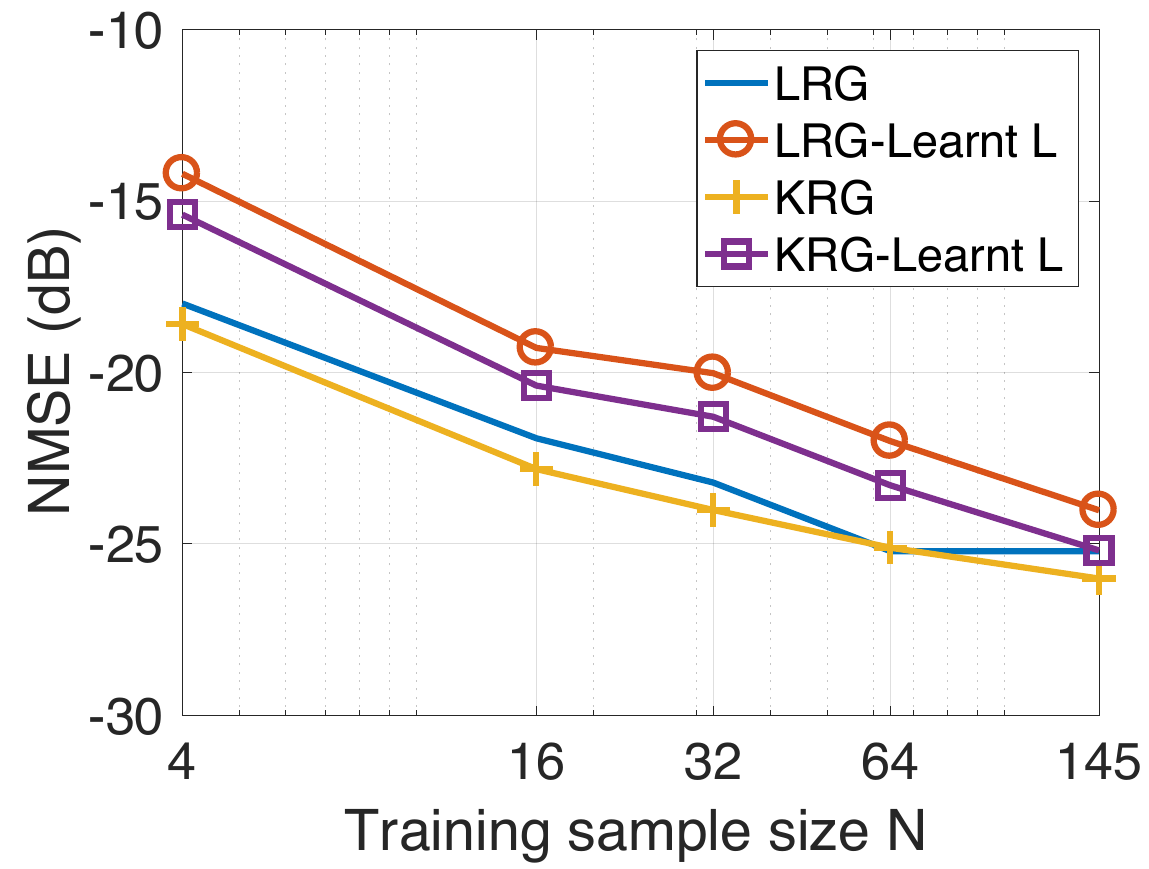}}\hspace{-.0in}\\
\subfigure[\hspace{-.2in}]{\includegraphics[width=2.4in]{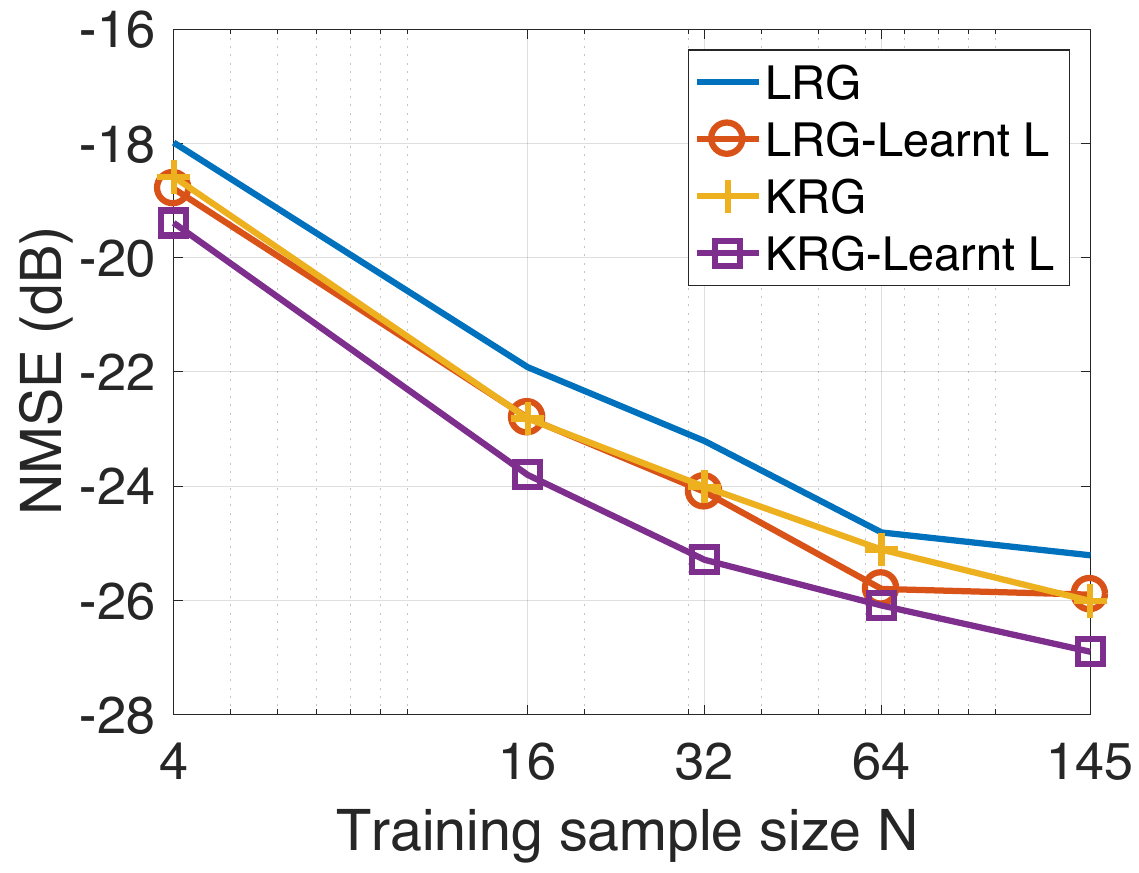}}\hspace{-.0in}
	\end{array}
	$
	\caption{{\color{black} Results for (D3) with graph learning for the fMRI data, when the training data samples are corrupted with additive white Gaussian noise at $10$dB SNR-level for $N=45$. (a) NMSE for the test data when initialized with $\mathbf{L}=\mathbf{0}$, and (b) when initialized with $\mathbf{L}$ corresponding to the graph based on atlas template\cite{Behjat_1}. %Geodesic Laplacian, (c) Estimated Laplacian from LRG, (d) Estimated Laplacian from KRG.
		% and (e) vectorized entries of the different Laplacians. The Laplacians have been scaled to have maximum eigenvalue equal to unity.
	}
	}
	\label{cere_learn}
\end{figure}
\begin{table*}
\caption{{\color{black}Comparison of the proposed methods and KRR for the experiment D3}}
	\centering
	\begin{tabular}{|c|c|c|c|c|c|c|c|}
		\hline
		
		\multicolumn{1}{|c|}{} &
		\multicolumn{3}{c|} {Kernel Ridge Regression (KRR)} &
		\multicolumn{4}{c|}{Proposed Methods} \\ \cline{2-8}	
		
		Training sample& Diffusion & Covariance kernel & Covariance kernel&LRG at &KRG at &LRG at &KRG at \\
		size $N$& kernel&at 10dB SNR& at 0dB SNR&10dB SNR&10dB SNR&0dB SNR&0dB SNR\\
		\hline
	%	5& 1.5$\times 10^{-5}$& 3.0& 5.8&-18.0&-18.2&-12.9&-13.9\\
	%	10&''&2.2&4.0&-21.5&-22.0&-16.2&-17.0\\
	%	15&''&0.8&1.9&-23.3&-24.1&-18.8&-19.6\\
	%	25 &''&-0.6&1.4&-25.2&-25.5&-19.5&-20.2\\
		145&0.01&-1.0&-0.7&-25.3&-25.7&-22.2&-23.5\\
		\hline
	\end{tabular}
	\label{tab:krr_cere}
\end{table*}

Finally, we consider the prediction of voxel intensities in the functional magnetic resonance imaging (fMRI) data obtained for the cerebellum region of the brain. { We apply our approach to predict the intensities at some of the voxels of the MRI, given the intensities at the other voxels. As before, we consider the training samples to be  corrupted with additive white Gaussian noise. 
	%{\color{black}Our hypothesis is the same as with our previous experiments -- that LRG and KRG outperform their conventional versions when the training data is limited and noisy. We also compare the performance of our methods with KRR. Further, we also investigate the case when an underlying graph is not known a-priori and is learnt from data using the method described in Section \ref{learngraph}. }  	
In the beginning, the graph is constructed from the voxels at the different slices of the MRI scans connected together to form a composite graph as shown in Figure \ref{cere}(a). The details of the image acquisition and the dataset may be found publicly at https://openfmri.org/dataset/ds000102. Each voxel is considered as a node of the graph and the voxel intensity to be the signal. The full data graph is of dimension 4000 obtained by mapping the 4000 cerebellum voxels anatomically following the atlas template \cite{cerebellum_atlas}. We refer the reader to \cite{Behjat_cerebellum} for further details on the construction of the voxel graph.  In our analysis, we consider only the first 100 voxels from the 4000 voxels to construct the dataset in our experiments. This is to reduce the computational complexity in performing the experiments. We use the intensity values at 10 of the voxels from the first slice as the input $\mathbf{x}\in\mathbb{R}^{10}$ to make predictions for the output $\mathbf{t}\in\mathbb{R}^{90}$ comprising the voxel instensities at 90 voxels present in the first and second slice. In Figure \ref{cere}(b), we show an instance of the voxel intensity signal over the full 4000 voxel graph. In Figures \ref{cere}(c) and (d), we show the corresponding input and output signals used in our experiments.} The dataset consists of 290 input-target data points or observation pairs. We use one half of the data for training and the other half for testing. We construct noisy training data at SNR levels of 10 dB and 0 dB. The NMSE is obtained by computing the average over 100 different random partitions of the entire data into the training and test sets. The results are shown in Figure \ref{cere}(e)-(f). We observe that LRG and KRG have a superior performance over their conventional counterparts, particularly for small $N$ and at low SNR-levels. 
 {\color{black}
 	
 	The performance of our methods and that of KRR with the maximum number of training samples is reported in Table \ref{tab:krr_cere}. We observe that KRR performs poorly in comparison with our approaches. The poor performance of KRR may be attributed to the relatively small number of samples available for reconstruction:  only 10\% of the total number of nodes are observed. Further, we note that KRR does not explicitly employ training data other than that used in the construction of the covariance kernel. The covariance kernel approach in turn also requires sufficient number of samples for a reliable reconstruction, which is not the case in our experiments owing to the adverse training data. All these factors explain why KRR performs rather poorly in this experiment.
}

{\color{black}We now consider the experiment with the learning of an underlying graph. We observe from Figure \ref{cere_learn}(a) that when initialized with the zero graph, the prediction performance of our method is comparable to that obtained  using the fixed atlas-template graph (given graph). We further consider the case when the graph learning iterations are initalized with $\mathbf{L}$ corresponding to the atlas-template. This is motivated by the observation that in many applications such as biomedical data, there is no single graph which is guaranteed to work the best for prediction. The goal of this experiment is then to investigate if a graph better suited to the regression task could be learnt starting from an existing non-trivial graph. 
	The prediction NMSE obtained for the test data in this case is shown in Figure \ref{cere_learn}(b). We observe that our method learns a graph better suited to the prediction in terms of the NMSE. This in turn shows that for this dataset, it is more suitable to jointly learn an underlying graph.}

\section{Conclusions}
\textcolor{black}{
We proposed a kernel regression method for predicting graph signal outputs from inputs that are not necessarily lying over a graph or for inputs agnostic to a graph. The resulting problem was shown to be a convex one resulting in an analytically tractable solution. Our approach presents a generalization of the standard kernel regression for graph signals. Experiments with synthesized and real-world graph signal datasets demonstrated the merit of our approach, particularly in the adverse scenarios of training with noise and limited datasizes. Our approach was shown to outperform a state-of-the-art method in real-world graph signal reconstruction problems. We further showed that our approach is also applicable in cases where an underlying graph is simultaneously estimated from the training data. 
}

	\section{Reproducible research}
	In the spirit of reproducible research, all the codes relevant to the experiments in this article are made available at https://www.researchgate.net/profile/Arun\textunderscore Venkitaraman and https://www.kth.se/ise/research/reproducibleresearch-1.433797.
	{\color{black}
	


\begin{thebibliography}{100}
\providecommand{\url}[1]{#1}
\csname url@samestyle\endcsname
\providecommand{\newblock}{\relax}
\providecommand{\bibinfo}[2]{#2}
\providecommand{\BIBentrySTDinterwordspacing}{\spaceskip=0pt\relax}
\providecommand{\BIBentryALTinterwordstretchfactor}{4}
\providecommand{\BIBentryALTinterwordspacing}{\spaceskip=\fontdimen2\font plus
\BIBentryALTinterwordstretchfactor\fontdimen3\font minus
  \fontdimen4\font\relax}
\providecommand{\BIBforeignlanguage}[2]{{%
\expandafter\ifx\csname l@#1\endcsname\relax
\typeout{** WARNING: IEEEtran.bst: No hyphenation pattern has been}%
\typeout{** loaded for the language `#1'. Using the pattern for}%
\typeout{** the default language instead.}%
\else
\language=\csname l@#1\endcsname
\fi
#2}}
\providecommand{\BIBdecl}{\relax}
\BIBdecl

\bibitem{Shuman}
D.~I. Shuman, S.~Narang, P.~Frossard, A.~Ortega, and P.~Vandergheynst, ``The
  emerging field of signal processing on graphs: Extending high-dimensional
  data analysis to networks and other irregular domains,'' \emph{IEEE Signal
  Process. Mag.}, vol.~30, no.~3, pp. 83--98, 2013.

\bibitem{Sandry1}
A.~Sandryhaila and J.~M.~F. Moura, ``Discrete signal processing on graphs,''
  \emph{IEEE Trans. Signal Process.}, vol.~61, no.~7, pp. 1644--1656, 2013.

\bibitem{Sandry2}
------, ``Big data analysis with signal processing on graphs: Representation
  and processing of massive data sets with irregular structure,'' \emph{IEEE
  Signal Process. Mag.}, vol.~31, no.~5, pp. 80--90, 2014.

\bibitem{Sandry3}
------, ``Discrete signal processing on graphs: Frequency analysis,''
  \emph{IEEE Trans. Signal Process.}, vol.~62, no.~12, pp. 3042--3054, 2014.

\bibitem{windowedGFT}
D.~I. Shuman, B.~Ricaud, and P.~Vandergheynst, ``A windowed graph {F}ourier
  transform,'' \emph{{IEEE} Statist. Signal Process. Workshop (SSP)}, pp.
  133--136, Aug 2012.

\bibitem{Narang2010}
S.~K. Narang and A.~Ortega, ``Local two-channel critically sampled filter-banks
  on graphs,'' \emph{Proc. IEEE Int. Conf. Image Process. (ICIP)}, pp.
  333--336, 2010.

\bibitem{Narang2012}
------, ``Perfect reconstruction two-channel wavelet filter banks for graph
  structured data,'' \emph{IEEE Trans. Signal Process.}, vol.~60, no.~6, pp.
  2786--2799, 2012.

\bibitem{Narang2013}
------, ``Compact support biorthogonal wavelet filterbanks for arbitrary
  undirected graphs,'' \emph{{IEEE} Trans. Signal Process.}, vol.~61, no.~19,
  pp. 4673--4685, 2013.

\bibitem{Coifman2006}
R.~R. Coifman and M.~Maggioni, ``Diffusion wavelets,'' \emph{Appl. Computat.
  Harmonic Anal.}, vol.~21, no.~1, pp. 53--94, 2006.

\bibitem{Ganesan}
D.~Ganesan, B.~Greenstein, D.~Estrin, J.~Heidemann, and R.~Govindan,
  ``Multiresolution storage and search in sensor networks,'' \emph{ACM Trans.
  Storage}, vol.~1, no.~3, pp. 277--315, 2005.

\bibitem{Hammond2011}
D.~K. Hammond, P.~Vandergheynst, and R.~Gribonval, ``Wavelets on graphs via
  spectral graph theory,'' \emph{Appl. Computat. Harmonic Anal.}, vol.~30,
  no.~2, pp. 129--150, 2011.

\bibitem{Wagner2}
R.~Wagner, V.~Delouille, and R.~Baraniuk, ``Distributed wavelet de-noising for
  sensor networks,'' \emph{Proc. 45th IEEE Conf. Decision Control}, pp.
  373--379, 2006.

\bibitem{vertexfreq}
D.~I. Shuman, B.~Ricaud, and P.~Vandergheynst, ``Vertex-frequency analysis on
  graphs,'' \emph{Appl. Comput. Harmonic Anal.}, vol.~40, no.~2, pp. 260 --
  291, 2016.

\bibitem{pyramidgraph}
D.~I. Shuman, M.~J. Faraji, and P.~Vandergheynst, ``A multiscale pyramid
  transform for graph signals,'' \emph{IEEE Trans. Signal Process.}, vol.~64,
  no.~8, pp. 2119--2134, April 2016.

\bibitem{pp_graph1}
O.~Teke and P.~P. Vaidyanathan, ``Extending classical multirate signal
  processing theory to graphs-part {I}: Fundamentals,'' \emph{{IEEE} Trans.
  Signal Process.}, vol.~65, no.~2, pp. 409--422, Jan 2017.

\bibitem{pp_graph2}
------, ``Extending classical multirate signal processing theory to graphs-part
  II: M-channel filter banks,'' \emph{IEEE Trans. Signal Process.}, vol.~65,
  no.~2, pp. 423--437, Jan 2017.

\bibitem{ArunSampta15}
\BIBentryALTinterwordspacing
A.~Venkitaraman, S.~Chatterjee, and P.~Handel, ``On {H}ilbert transform of
  signals on graphs,'' \emph{Proc. Sampling Theory Appl.}, 2015. [Online].
  Available: \url{http://w.american.edu/cas/sampta/papers/a13-venkitaraman.pdf}
\BIBentrySTDinterwordspacing

\bibitem{ArunGHT}
A.~Venkitaraman, S.~Chatterjee, and P.~H{\"{a}}ndel, ``On {H}ilbert transform,
  analytic signal, and modulation analysis for signals over graphs,''
  \emph{Signal Process.}, vol. 156, pp. 106--115, 2019.

\bibitem{Tremblay}
N.~Tremblay, G.~Puy, P.~Borgnat, R.~Gribonval, and P.~Vandergheynst,
  ``{Accelerated spectral clustering using graph filtering of random
  signals},'' \emph{{ IEEE Int. Conf. Acoust. Speech Signal Process.}}, 2016.

\bibitem{Tremblay2}
N.~Tremblay and P.~Borgnat, ``Joint filtering of graph and graph-signals,''
  \emph{Asilomar Conf. Signals Syst. Comput.}, pp. 1824--1828, Nov 2015.

\bibitem{chen2}
S.~Chen, R.~Varma, A.~Sandryhaila, and J.~Kovacevic, ``Discrete signal
  processing on graphs: Sampling theory,'' \emph{{IEEE} Trans. Signal
  Process.}, vol.~63, no.~24, pp. 6510--6523, Dec 2015.

\bibitem{graphsamp1}
F.~Gama, A.~G. Marques, G.~Mateos, and A.~Ribeiro, ``Rethinking sketching as
  sampling: Linear transforms of graph signals,'' \emph{Proc. Asilomar Conf.
  Signals Syst. Comput.}, pp. 522--526, Nov 2016.

\bibitem{graphsamp2}
S.~P. Chepuri and G.~Leus, ``Subsampling for graph power spectrum estimation,''
  \emph{Proc. IEEE Sensor Array Multichannel Signal Process. Workshop (SAM)},
  pp. 1--5, July 2016.

\bibitem{graphsamp3}
S.~Chen, R.~Varma, A.~Singh, and J.~Kova{\v c}evi{\'c}, ``Signal recovery on
  graphs: Fundamental limits of sampling strategies,'' \emph{IEEE Trans. Signal
  Inf. Process. Netw.}, vol.~2, no.~4, pp. 539--554, Dec 2016.

\bibitem{graphsamp4}
A.~G. Marques, S.~Segarra, G.~Leus, and A.~Ribeiro, ``Sampling of graph signals
  with successive local aggregations,'' \emph{{IEEE} Trans. Signal Process.},
  vol.~64, no.~7, pp. 1832--1843, April 2016.

\bibitem{graphsamp5}
H.~Q. Nguyen and M.~N. Do, ``Downsampling of signals on graphs via maximum
  spanning trees,'' \emph{{IEEE} Trans. Signal Process.}, vol.~63, no.~1, pp.
  182--191, Jan 2015.

\bibitem{graphsamp6}
M.~Tsitsvero, S.~Barbarossa, and P.~D. Lorenzo, ``Signals on graphs:
  Uncertainty principle and sampling,'' \emph{{IEEE} Trans. Signal Process.},
  vol.~64, no.~18, pp. 4845--4860, 2016.

\bibitem{anis}
A.~Anis, A.~Gadde, and A.~Ortega, ``Efficient sampling set selection for
  bandlimited graph signals using graph spectral proxies,'' \emph{{IEEE} Trans.
  Signal Process.}, vol.~64, no.~14, pp. 3775--3789, July 2016.

\bibitem{graphsamp10}
L.~F.~O. {Chamon} and A.~{Ribeiro}, ``Greedy sampling of graph signals,''
  \emph{IEEE Trans. Sig. Process.}, vol.~66, no.~1, pp. 34--47, Jan 2018.

\bibitem{graphsamp11}
S.~P. {Chepuri} and G.~{Leus}, ``Graph sampling for covariance estimation,''
  \emph{IEEE Trans. Signal Inf. Process. Netw.}, vol.~3, no.~3, pp. 451--466,
  Sep. 2017.

\bibitem{KRG_R1}
S.~K. Narang, A.~Gadde, and A.~Ortega, ``Signal processing techniques for
  interpolation in graph structured data,'' in \emph{IEEE Int. Conf. Acoust.
  Speech Signal Process.}, May 2013, pp. 5445--5449.

\bibitem{graphPCA1}
N.~Shahid, V.~Kalofolias, X.~Bresson, M.~Bronstein, and P.~Vandergheynst,
  ``Robust principal component analysis on graphs,'' \emph{IEEE Int. Conf.
  Comput. Vision (ICCV)}, pp. 2812 -- 2820, 2015.

\bibitem{graphPCA2}
N.~Shahid, N.~Perraudin, V.~Kalofolias, G.~Puy, and P.~Vandergheynst, ``Fast
  robust pca on graphs,'' \emph{IEEE J. Selected Topics Signal Process.},
  vol.~10, no.~4, pp. 740--756, June 2016.

\bibitem{Thanou2014}
D.~Thanou, {D. I Shuman}, and P.~Frossard, ``Learning parametric dictionaries
  for signals on graphs,'' \emph{IEEE Trans. Signal Process.}, vol.~62, no.~15,
  pp. 3849--3862, 2014.

\bibitem{graphdict1}
D.~Thanou and P.~Frossard, ``Multi-graph learning of spectral graph
  dictionaries,'' in \emph{IEEE Int. Conf. Acoust. Speech Signal Process.},
  April 2015, pp. 3397--3401.

\bibitem{graphdict3}
D.~Thanou, D.~I. Shuman, and P.~Frossard, ``Learning parametric dictionaries
  for signals on graphs,'' \emph{IEEE Trans. Signal Process.}, vol.~62, no.~15,
  pp. 3849--3862, Aug 2014.

\bibitem{dualgraphelad}
Y.~Yankelevsky and M.~Elad, ``Dual graph regularized dictionary learning,''
  \emph{IEEE Trans. Signal Inf. Process. Netw.}, vol.~2, no.~4, pp. 611--624,
  Dec 2016.

\bibitem{statgraph1}
B.~Girault, ``Stationary graph signals using an isometric graph translation,''
  \emph{Proc. Eur. Signal Process. Conf. (EUSIPCO)}, pp. 1516--1520, Aug 2015.

\bibitem{statgraph2}
S.~Segarra, A.~G. Marques, G.~Leus, and A.~Ribeiro, ``Stationary graph
  processes: Nonparametric spectral estimation,'' \emph{Proc. IEEE Sensor Array
  Multichannel Signal Process. Workshop (SAM)}, pp. 1--5, July 2016.

\bibitem{statgraph2_journal}
A.~G. {Marques}, S.~{Segarra}, G.~{Leus}, and A.~{Ribeiro}, ``Stationary graph
  processes and spectral estimation,'' \emph{IEEE Trans. Signal Process.},
  vol.~65, no.~22, pp. 5911--5926, Nov 2017.

\bibitem{statgraph3}
N.~Perraudin and P.~Vandergheynst, ``Stationary signal processing on graphs,''
  \emph{IEEE Trans. Sig. Proc.}, vol.~65, no.~13, pp. 3462--3477, Jul. 2017.

\bibitem{statgraph4}
B.~Girault, P.~Goncalves, and E.~Fleury, ``Translation on graphs: An isometric
  shift operator,'' \emph{IEEE Signal Process. Lett.}, vol.~22, no.~12, pp.
  2416--2420, Dec. 2015.

\bibitem{statgraph5}
B.~Girault, ``Stationary graph signals using an isometric graph translation,''
  \emph{Proc. Eur. Signal Process. Conf. (EUSIPCO)}, 2015.

\bibitem{Berger17}
P.~Berger, G.~Hannak, and G.~Matz, ``Graph signal recovery via primal-dual
  algorithms for total variation minimization,'' \emph{IEEE J. Selected Topics
  Signal Process.}, vol.~11, no.~6, pp. 842--855, Sept 2017.

\bibitem{chen1}
S.~Chen, A.~Sandryhaila, J.~M.~F. Moura, and J.~Kovacevic, ``Signal recovery on
  graphs: Variation minimization,'' \emph{{IEEE} Trans. Signal Process.},
  vol.~63, no.~17, pp. 4609--4624, Sept 2015.

\bibitem{KRG_R2}
X.~Wang, M.~Wang, and Y.~Gu, ``A distributed tracking algorithm for
  reconstruction of graph signals,'' \emph{IEEE J. Selected Topics Signal
  Process.}, vol.~9, no.~4, pp. 728--740, June 2015.

\bibitem{KRG_R3}
P.~D. Lorenzo, S.~Barbarossa, P.~Banelli, and S.~Sardellitti, ``Adaptive least
  mean squares estimation of graph signals,'' \emph{IEEE Trans. Signal Inf.
  Process. Netw.}, vol.~2, no.~4, pp. 555--568, Dec 2016.

\bibitem{Dong:2016}
X.~Dong, D.~Thanou, P.~Frossard, and P.~Vandergheynst, ``Learning {L}aplacian
  matrix in smooth graph signal representations,'' \emph{IEEE Trans. Signal
  Process.}, vol.~64, no.~23, pp. 6160--6173, Dec. 2016.

\bibitem{graphlearn8}
Y.~Shen, B.~Baingana, and G.~B. Giannakis, ``Tensor decompositions for
  identifying directed graph topologies and tracking dynamic networks,''
  \emph{IEEE Trans. Signal Process.}, vol.~65, no.~14, pp. 3675--3687, July
  2017.

\bibitem{graphlearn2}
C.~Hu, L.~Cheng, J.~Sepulcre, G.~E. Fakhri, Y.~M. Lu, and Q.~Li, ``A graph
  theoretical regression model for brain connectivity learning of {A}lzheimer's
  disease,'' \emph{Proc. IEEE Int. Symp. Biomed. Imag.}, pp. 616--619, April
  2013.

\bibitem{graphlearn3}
S.~I. Daitch, J.~A. Kelner, and D.~A. Spielman, ``Fitting a graph to vector
  data,'' \emph{Proc. Int. Conf. Mach. Learn.}, pp. 201--208, 2009.

\bibitem{graphlearn4}
N.~Leonardi and D.~{Van~De~Ville}, ``Wavelet frames on graphs defined by {fMRI}
  functional connectivity,'' in \emph{IEEE Int. Symp. Biomed. Imag.}, March
  2011, pp. 2136--2139.

\bibitem{graphlearn5}
N.~Leonardi, J.~Richiardi, M.~Gschwind, S.~Simioni, J.-M. Annoni, M.~Schluep,
  P.~Vuilleumier, and D.~{Van~De~Ville}, ``Principal components of functional
  connectivity: A new approach to study dynamic brain connectivity during
  rest,'' \emph{NeuroImage}, vol.~83, pp. 937--950, 2013.

\bibitem{graphlearn9}
S.~Segarra, A.~G. Marques, G.~Mateos, and A.~Ribeiro, ``Network topology
  inference from spectral templates,'' \emph{IEEE Trans. Signal Inf. Process.
  Netw.}, vol.~3, no.~3, pp. 467--483, Sept 2017.

\bibitem{Chepuri_laplacian}
S.~P. Chepuri, S.~Liu, G.~Leus, and A.~Hero, ``Learning sparse graphs under
  smoothness prior,'' \emph{Proc. IEEE Int. Conf. Acoust. Speech Signal
  Process.}, pp. 6508--6512, 2017.

\bibitem{SR7}
B.~Pasdeloup, V.~Gripon, G.~Mercier, D.~Pastor, and M.~G. Rabbat,
  ``Characterization and inference of graph diffusion processes from
  observations of stationary signals,'' \emph{IEEE Trans. Signal Inf. Process.
  Netw.}, pp. 1--1, 2017.

\bibitem{graphrecon}
S.~{Segarra}, A.~G. {Marques}, G.~{Leus}, and A.~{Ribeiro}, ``Reconstruction of
  graph signals through percolation from seeding nodes,'' \emph{IEEE Trans
  Signal Process.}, vol.~64, no.~16, pp. 4363--4378, Aug 2016.

\bibitem{gl_krg}
\BIBentryALTinterwordspacing
A.~Venkitaraman, H.~P. Maretic, S.~Chatterjee, and P.~Frossard, ``Supervised
  linear regression for graph learning from graph signals,'' \emph{{CoRR}},
  vol. abs/1811.01586, 2018. [Online]. Available:
  \url{http://arxiv.org/abs/1811.01586}
\BIBentrySTDinterwordspacing

\bibitem{gsp_overview_ortega}
A.~Ortega, P.~Frossard, J.~Kova{\v c}evi{\'c}, J.~M.~F. Moura, and
  P.~Vandergheynst, ``Graph signal processing: Overview, challenges, and
  applications,'' \emph{Proc. IEEE}, vol. 106, no.~5, pp. 808--828, May 2018.

\bibitem{SVM}
C.~Cortes and V.~Vapnik, ``Support-vector networks,'' \emph{J. Mach. Learn.},
  vol.~20, no.~3, pp. 273--297, 1995.

\bibitem{GP_Seeger}
M.~Seeger, ``Gaussian processes for machine learning,'' \emph{Int. J. Neural
  Syst.}, vol.~14, no.~02, pp. 69--106, 2004.

\bibitem{deeplearning}
L.~Yann, Y.~Bengio, , and G.~Hinton, ``Deep learning,'' \emph{Nature}, vol.
  521, no. 7553, pp. 436--444, 2015.

\bibitem{kernel_deeplearning}
Y.~Cho and L.~K. Saul, ``Kernel methods for deep learning,'' in \emph{Adv.
  Neural Inf. Process. Syst.}, Y.~Bengio, D.~Schuurmans, J.~D. Lafferty,
  C.~K.~I. Williams, and A.~Culotta, Eds.\hskip 1em plus 0.5em minus
  0.4em\relax Curran Associates, Inc., 2009, pp. 342--350.

\bibitem{ELM_kernel1}
G.-B. Huang, ``An insight into extreme learning machines: Random neurons,
  random features and kernels,'' \emph{Cogn. Computat.}, vol.~6, pp. 376--390,
  2014.

\bibitem{ELM_kernel2}
A.~Iosifidis, A.~Tefas, and I.~Pitas, ``On the kernel extreme learning machine
  classifier,'' \emph{Patt. Recognit. Lett.}, vol.~54, pp. 11 -- 17, 2015.

\bibitem{diffusionkernels}
R.~I. Kondor and J.~Lafferty, ``Diffusion kernels on graphs and other discrete
  structures,'' \emph{Proc. Int. Conf. Mach. Learn. {ICML}}, pp. 315--322,
  2002.

\bibitem{Smola2003}
A.~J. Smola and R.~Kondor, \emph{Kernels and Regularization on Graphs}.\hskip
  1em plus 0.5em minus 0.4em\relax Berlin, Heidelberg: Springer Berlin
  Heidelberg, 2003, pp. 144--158.

\bibitem{graphlabel2}
M.~Belkin, I.~Matveeva, and P.~Niyogi, ``Tikhonov regularization and
  semi-supervised learning on large graphs,'' \emph{Proc. IEEE Int. Conf.
  Acoust. Speech Signal Process.}, May 2004.

\bibitem{graphlabel3}
------, ``Regularization and semi-supervised learning on large graphs,'' in
  \emph{Proc. 17th Annual Conf. Learn. Theory}, J.~Shawe-Taylor and Y.~Singer,
  Eds.\hskip 1em plus 0.5em minus 0.4em\relax Berlin, Heidelberg: Springer
  Berlin Heidelberg, 2004, pp. 624--638.

\bibitem{graphlabel4}
A.~Argyriou, M.~Herbster, and M.~Pontil, ``Combining graph {L}aplacians for
  semi-supervised learning,'' \emph{Proc. Neural Info. Process. Syst.}, pp.
  67--74, 2005.

\bibitem{graphlabel5}
M.~Belkin, P.~Niyogi, and V.~Sindhwani, ``Manifold regularization: A geometric
  framework for learning from labeled and unlabeled examples,'' \emph{J. Mach.
  Learn. Res.}, vol.~7, pp. 2399--2434, Dec. 2006.

\bibitem{graphlabel6}
K.~Zhang, L.~Lan, J.~T. Kwok, S.~Vucetic, and B.~Parvin, ``Scaling up
  graph-based semisupervised learning via prototype vector machines,''
  \emph{IEEE Trans. Neural Netw. Learn. Syst.}, vol.~26, no.~3, pp. 444--457,
  March 2015.

\bibitem{graphlabel7}
K.~I. Kim, J.~Tompkin, H.~Pfister, and C.~Theobalt, ``Context-guided diffusion
  for label propagation on graphs,'' in \emph{IEEE Int. Conf. Comput. Vision
  (ICCV)}, Dec 2015, pp. 2776--2784.

\bibitem{takeda2008deblurring}
H.~Takeda, S.~Farsiu, and P.~Milanfar, ``Deblurring using regularized locally
  adaptive kernel regression,'' \emph{IEEE Trans. Image Process.}, vol.~17,
  no.~4, pp. 550--563, 2008.

\bibitem{dou2017object}
H.~Dou, D.~Ming, Z.~Yang, Z.~Pan, Y.~Li, and J.~Tian, ``Object-based visual
  saliency via {L}aplacian regularized kernel regression,'' \emph{IEEE Trans.
  Multimedia}, vol.~19, no.~8, pp. 1718--1729, 2017.

\bibitem{kerbrain_5}
M.~K. Chung and J.~Taylor, ``Diffusion smoothing on brain surface via finite
  element method,'' \emph{Proc. IEEE Int. Symp. Biomed. Imag.}, pp. 432--435,
  2004.

\bibitem{kerbrain_1}
M.~K. Chung, P.~Bubenik, and P.~T. Kim, ``Persistence diagrams of cortical
  surface data,'' \emph{Int. Conf. Inf. Proc. Med. Imag.}, pp. 386--397, 2009.

\bibitem{kerbrain_6}
S.~Seo, M.~K. Chung, and H.~K. Vorperian, ``Heat kernel smoothing using
  {L}aplace-{B}eltrami eigenfunctions,'' \emph{Proc. Med. Image. Comput.
  Assist. Interv.}, pp. 505--512, 2010.

\bibitem{kerbrain_2}
D.~{Pachauri}, C.~{Hinrichs}, M.~K. {Chung}, S.~C. {Johnson}, and V.~{Singh},
  ``Topology-based kernels with application to inference problems in
  {A}lzheimer's disease,'' \emph{IEEE Trans. Med. Imag.}, vol.~30, no.~10, pp.
  1760--1770, Oct 2011.

\bibitem{kerbrain_3}
J.~Reininghaus, S.~Huber, U.~Bauer, and R.~Kwitt, ``A stable multi-scale kernel
  for topological machine learning,'' \emph{Proc. IEEE Conf. Comput. Vis.
  Pattern Recognit. (CVPR)}, pp. 4741--4748, Jun. 2015.

\bibitem{kerbrain_0}
V.~{Solo}, J.~{Poline}, M.~A. {Lindquist}, S.~L. {Simpson}, F.~D. {Bowman},
  M.~K. {Chung}, and B.~{Cassidy}, ``Connectivity in {fMRI}: Blind spots and
  breakthroughs,'' \emph{IEEE Trans. Med. Imag.}, vol.~37, no.~7, pp.
  1537--1550, July 2018.

\bibitem{kergraph1}
D.~Romero, M.~Ma, and G.~B. Giannakis, ``Kernel-based reconstruction of graph
  signals,'' \emph{{IEEE} Trans. Signal Process.}, vol.~65, no.~3, pp.
  764--778, Feb 2017.

\bibitem{kergraph2}
------, ``Estimating signals over graphs via multi-kernel learning,'' in
  \emph{{IEEE} Statist. Signal Process. Workshop (SSP)}, June 2016, pp. 1--5.

\bibitem{kergraph3}
V.~N. Ioannidis, D.~Romero, and G.~B. Giannakis, ``Kernel-based reconstruction
  of space-time functions via extended graphs,'' in \emph{Asilomar Conf.
  Signals Syst. Comput.}, Nov 2016, pp. 1829--1833.

\bibitem{kergraph4}
D.~{Romero}, V.~N. {Ioannidis}, and G.~B. {Giannakis}, ``Kernel-based
  reconstruction of space-time functions on dynamic graphs,'' \emph{IEEE J.
  Select. Topics Signal Process.}, vol.~11, no.~6, pp. 856--869, Sep. 2017.

\bibitem{IOANNIDIS2018173}
V.~N. Ioannidis, M.~Ma, A.~N. Nikolakopoulos, G.~B. Giannakis, and D.~Romero,
  \emph{Kernel-Based Inference of Functions Over Graphs}.\hskip 1em plus 0.5em
  minus 0.4em\relax Butterworth-Heinemann, 2018, pp. 173 -- 198.

\bibitem{CHUNG201563}
M.~K. Chung, A.~Qiu, S.~Seo, and H.~K. Vorperian, ``Unified heat kernel
  regression for diffusion, kernel smoothing and wavelets on manifolds and its
  application to mandible growth modeling in {CT} images,'' \emph{Med. Image
  Anal.}, vol.~22, no.~1, pp. 63 -- 76, 2015.

\bibitem{baingana1}
Y.~Shen, B.~Baingana, and G.~B. Giannakis, ``Kernel-based structural equation
  models for topology identification of directed networks,'' \emph{{IEEE}
  Trans. Signal Process.}, vol.~65, no.~10, pp. 2503--2516, May 2017.

\bibitem{Arun_GPG}
A.~Venkitaraman, S.~Chatterjee, and P.~H{\"a}ndel, ``Gaussian processes over
  graphs,'' \emph{{CoRR} https://arxiv.org/abs/1803.05776}, vol.
  abs/1803.05776, 2018.

\bibitem{Bishop}
C.~M. Bishop, \emph{Pattern Recognition and Machine Learning (Information
  Science and Statistics)}.\hskip 1em plus 0.5em minus 0.4em\relax Secaucus,
  NJ, USA: Springer-Verlag New York, Inc., 2006.

\bibitem{Loan1}
C.~F.~V. Loan, ``The ubiquitous {K}ronecker product,'' \emph{J. Comput. Appl.
  Math.}, vol. 123, no. 1--2, pp. 85--100, 2000.

\bibitem{RasmussenGP}
C.~E. Rasmussen and C.~K.~I. Williams, \emph{Gaussian Processes for Machine
  Learning (Adaptive Computation and Machine Learning)}.\hskip 1em plus 0.5em
  minus 0.4em\relax The MIT Press, 2005.

\bibitem{graphlearn6}
P.~K. Shivaswamy and T.~Jebara, \emph{{L}aplacian Spectrum Learning}.\hskip 1em
  plus 0.5em minus 0.4em\relax Berlin, Heidelberg: Springer Berlin Heidelberg,
  2010, pp. 261--276.

\bibitem{graphlearn7}
X.~Dong, D.~Thanou, P.~Frossard, and P.~Vandergheynst, ``Laplacian matrix
  learning for smooth graph signal representation,'' in \emph{IEEE Int. Conf.
  Acoust. Speech Signal Process.}, April 2015, pp. 3736--3740.

\bibitem{kalofolias16}
V.~Kalofolias, ``How to learn a graph from smooth signals,'' \emph{Proc. Int.
  Conf. Artif. Intell. Statist.}, vol.~51, pp. 920--929, May 2016.

\bibitem{Chung}
F.~R.~K. Chung, \emph{Spectral Graph Theory}.\hskip 1em plus 0.5em minus
  0.4em\relax AMS, 1996.

\bibitem{multikernel_5}
C.~Cortes, M.~Mohri, and A.~Rostamizadeh, ``L2 regularization for learning
  kernels,'' in \emph{Proc. Conf. Uncertainty in Artif. Intell.}, ser. UAI '09,
  2009, pp. 109--116.

\bibitem{multikernel_Arun}
A.~{Venkitaraman}, S.~{Chatterjee}, and P.~{H{\"a}ndel}, ``Multi-kernel
  regression for graph signal processing,'' in \emph{IEEE Int. Conf. Acoust.
  Speech Signal Process. (ICASSP)}, April 2018, pp. 4644--4648.

\bibitem{Newman}
M.~E.~J. Newman, \emph{Networks: An Introduction}.\hskip 1em plus 0.5em minus
  0.4em\relax Oxford University Press, 2010.

\bibitem{SMHI}
\BIBentryALTinterwordspacing
Swedish meteorological and hydrological institute (smhi). [Online]. Available:
  \url{http://opendata-download-metobs.smhi.se/}
\BIBentrySTDinterwordspacing

\bibitem{Behjat_1}
H.~Behjat, U.~Richter, D.~{Van~De~Ville}, and L.~S{\"o}rnmo, ``Signal-adapted
  tight frames on graphs,'' \emph{{IEEE} Trans. Signal Process.}, vol.~64,
  no.~22, pp. 6017--6029, Nov 2016.

\bibitem{cerebellum_atlas}
J.~Diedrichsen, J.~H. Balsters, J.~Flavell, E.~Cussans, and N.~Ramnani, ``A
  probabilistic {MR} atlas of the human cerebellum,'' \emph{NeuroImage},
  vol.~46, no.~1, pp. 39 -- 46, 2009.

\bibitem{Behjat_cerebellum}
H.~Behjat, N.~Leonardi, L.~S{\"o}rnmo, and D.~{Van~De~Ville},
  ``Anatomically-adapted graph wavelets for improved group-level {fMRI}
  activation mapping,'' \emph{NeuroImage}, vol. 123, pp. 185 -- 199, 2015.

\end{thebibliography}
\end{document}